\documentclass[5p]{elsarticle}
\journal{Journal of Computational Physics}

\usepackage{geometry} % decreases margins
\usepackage{subcaption}
\usepackage[T1]{fontenc}
\usepackage{amsmath,amsfonts,amssymb}
\usepackage{wrapfig}	% wrap text around figures; https://en.wikibooks.org/wiki/LaTeX/Floats,_Figures_and_Captions
\usepackage{todonotes}	% \todo{}: side notes at the margins; \todo[inline]{}: inline notes; \missingfigure{}; \listoftodos
\usepackage{color, soul}	% Highlighting, use \hl{...}
\usepackage{verbatim}	% comment out chunks of text
\usepackage{pifont} % Provides the \ding command
\usepackage{hyperref}

\newcommand{\orcid}[1]{\href{#1}{\includegraphics[scale=0.035]{figures/orcid.png}}}

\definecolor{boxback}{HTML}{DAE5F0}
\definecolor{boxframe}{HTML}{0F365B}
{
\end{tcolorbox}   
}

\def\code#1{\texttt{#1}}

\begin{document}
\begin{frontmatter}
\title{Spectral Learning of Magnetized Plasma Dynamics: A Neural Operator Application}
\author[1]{Roberta Duarte\corref{cor1}}
\ead{roberta.pereira@usp.br}
\author[1,2]{Rodrigo Nemmen}
\ead{rodrigo.nemmen@iag.usp.br}
\author[1]{Reinaldo Santos-Lima}
\ead{reinaldo.lima@iag.usp.br}
\cortext[cor1]{Corresponding author}
\affiliation[1]{organization={Instituto de Astronomia, Geofísica e Ciências Atmosféricas, Universidade de São Paulo},
	addressline={Rua do Matão 1226},
	postcode={05508-090},
	city={São Paulo, São Paulo},
	country={Brazil}}
\affiliation[2]{organization={Kavli Institute for Particle Astrophysics and Cosmology, Stanford University},
	addressline={452 Lomita Mall},
	postcode={94305},
	city={Stanford, California},
	country={United States of America}}

\begin{abstract}
Fourier neural operators (FNOs) provide a mesh-independent way to learn solution operators for partial differential equations, yet their efficacy for magnetized turbulence is largely unexplored. Here we train an FNO surrogate for the 2-D Orszag–Tang vortex, a canonical non-ideal magnetohydrodynamic (MHD) benchmark, across an ensemble of viscosities and magnetic diffusivities. On unseen parameter settings the model achieves a mean-squared error of  $\approx 6\times10^{-3}$ in velocity and $\approx 10^{-3}$ in magnetic field, reproduces energy spectra and dissipation rates within $96\%$ accuracy, and retains temporal coherence over long timescales. Spectral analysis shows accurate recovery of large- and intermediate-scale structures, with degradation at the smallest resolved scales due to Fourier-mode truncation. Relative to a UNet baseline the FNO cuts error by $97\%$, and compared with a high-order finite-volume solver it delivers a $25\times$ inference speed-up, offering a practical path to rapid parameter sweeps in MHD simulations.
\end{abstract}

%\begin{highlights}
%	\item Neural networks embedded in PDEs can violate physical bounds unless constrained.
%	\item Motivated by established schemes, a new neural network approach is introduced.
%	\item It is demonstrated for subgrid-scale modeling for turbulent combustion.
%\end{highlights}

\end{frontmatter}

\section{Introduction} \label{sec:intro}

The application of machine learning (ML) to simulate physical problems has gained attention in recent years \citep{cranmer2020, greydanus2019, Pfaff2020}. Researchers have explored ML techniques in multiple domains, including discrete and continuous chaotic systems \citep{Pathak2018a, Mohan2019, Breen2020}. One particular noteworthy application has been fluid mechanics.

Fluid mechanics as a field is going through a rennassaince from the application of machine learning methods that make use of the large volumes of data coming from laboratory experiments, observations, and numerical simulations \citep{Brunton2020}. Here, we briefly review some of the advancements in computational fluid dynamics due to ML, with a focus on deep learning and solution operator methods.

%\todo[inline, color=yellow!20]{RN: The previous literature review on ML for CFD is incomplete. I added a more complete revision, taking into account what you wrote before. I divided the literature review in three parts:
%1. Deep learning with CNNs
%2. PINNs
%3. Solution operator learning}
% DL
Many authors treated the fluid dynamics problem as a computer vision problem where the field values at a given time are analogous to an image, and trained convolutional neural networks (CNN) on simulation data \citep{Guo2016, Tompson2016, Kim2019, Kochkov2021, Duarte2022}. Some authors used a similar approach but chose other archictetures over CNN such as graph neural network  \citep{Pfaff2020} or LSTM \citep{Mohan2019} which in some cases demonstrated superior accuracy compared to traditional UNet architectures.

% PINNs
\cite{Raissi2019} introduced physics-informed neural networks (PINNs) as a general framework for solving partial differential equations (PDEs)---such as the ones that govern fluids---by incorporating physics in the loss function. Subsequently, other authors continued exploring PINNs specifically for fluids \citep{Mao2020, Jin2021, eivazi2024}. One conclusion from such works is that PINNs are not as accurate compared to numerical solvers for forward problems but they are superior for inverse problems, that is recovering high-fidelity flow data from sparse and noisy measurements.

% Solution operators
\cite{li2021,Kovachki2021} introduced the pioneering concept of neural operators as architectures that learn the solution operator to PDEs irrespective of mesh details. In particular, \cite{li2021} introduced the Fourier neural operator (FNO) which leverages the Fourier transform to represent the solution in the frequency domain. FNOs are able to capture long-range dependencies in the solution and the promise of mesh-free generalization. However, because the Fourier transform inherently assumes periodicity in the domain, FNOs shine for problems with periodic boundary conditions. 

% plasmas and magnetic fields
Despite the potential of FNOs, they remain relatively unexplored for plasmas, a condition that occurs when the gas temperature is high enough to ionize atoms and produce a gas full of free electrons, such that the collective, long-range electromagnetic interaction in
the gas is dominant over the individual, small-range electrostatic interactions between particles. Exploring and understanding the application of ML to plasmas is important because most of the matter in the universe is in that form. There are two main approaches to studying plasma dynamics \citep{Chen2016}. One is using kinetic theory and describing particle behavior using distribution functions and solving the Vlasov equation. The other is magnetohydrodynamics (MHD) which treats the plasma as a fluid and is appropriate for
large scale and low frequency phenomena, which is spatial (temporal) scales much larger (longer) than the kinetic scales, such as the ion skin depth (inverse of the ion plasma frequency) and ion Larmor radius (ion gyration time) in magnetized plasmas. MHD consists of a set of partial differential equations that describe the conservation of energy, momentum and mass coupled with Maxwell's equations which characterize the dynamics of the collective
electromagnetic field. There are many phenomena where plasmas play a crucial role and can be described with MHD, such as stellar atmospheres, the magnetospheres of stars, planets, black holes, gas accretion onto compact objects, the interstellar medium, and the intracluster medium of galaxies. \citep{Kulsrud2004, Meier2012}. 

% MHD equations and numerical solvers
There is a wide variety of codes available for this purpose, using different techniques such as finite volume  and spectral methods \citep{Mignone2007, Llambay2016, Stone2020}. 
% MHD benchmarks; Orszag-Tang vortex
The correctness and performance of numerical MHD solvers are evaluated by simulating classical and well-established problems. Some classic test problems include the Brio  $\&$  Wu shock tube test, magnetic loop advection, and current sheet evolution \citep{Mignone2007,Llambay2016}. Among these, the Orszag-Tang Vortex \citep{orszag1979} is a well-known and standard benchmark, often used to evaluate how effectively an MHD code generates turbulence, shocks and dissipation \citep{Parashar2010}.

In this work, we focus on MHD problems. \cite{Rosofsky2023} used an FNO-based method to learn 2D incompressible MHD simulations for a variety of initial conditions and and Reynolds numers, which are dimensionless quantities characterizing the ratio between the time scale of the momentum or magnetic dissipation via microphysical processes and the dynamical time scales of the problem. They found that for ${\rm Re} \leq 250$ their FNO method is quite accurate while for larger values it loses accuracy for longer simulation times. Here, we simulate the
Orszag-Tang vortex problem using the FNO method.

% this paper
We aim to evaluate how effectively FNO can learn and predict dynamics in a magnetized, turbulent environment. The structure of this work is as follows: $\S 2$ provides an overview of the Orszag-Tang vortex and its characteristics. In $\S 3$, we detail the methods employed, focusing on the FNO framework. Section $\S 4$ outlines the data generation and preparation process. The results are presented in $\S 5$, followed by an in-depth discussion in $\S 6$.

\section{Orszag-Tang Vortex}

The Orszag-Tang vortex problem was introduced by \cite{orszag1979}. The problem consists of initial conditions which generate a vortex and exhibit turbulent behavior as the system evolves. The vortex evolution shows interactions between shock waves with a range of speed regimes \citep{snow2021}. Several works studied the vortex deeply by analyzing the shock-shock interactions, MHD discontinuities, and turbulence \citep{Dahlburg1989, balbas2005, Uritsky2010}. Since these are characteristics of interest in the MHD study, the Orszag-Tang vortex is an ideal standard test to investigate how robust a numerical code is when these nonlinear phenomena arise. The setup is a 2D simulation with periodic boundaries where the magnetic field follows a sinusoidal profile perturbed by a sinusoidal velocity profile. The parameters of interest in this system are kinematic viscosity $(\nu)$ and Ohmic diffusivity $(\eta)$, as they directly influence energy dissipation. As the simulation progresses, vortices are formed, and there is the onset of chaotic, turbulent dynamics and shocks. These characteristics make the Orszag-Tang vortex an ideal test case for evaluating numerical solvers, hence our choice to use this well-known problem. 

The Orszag-Tang vortex is governed by the nondeal
MHD equations, a set of four PDEs representing conservation laws. The first of these is the continuity equation, given by:

\begin{equation}
    \frac{\partial \rho}{\partial t} + \nabla \cdot (\rho \mathbf{v}) = 0,
\end{equation}

\noindent where $\rho$ is the density and $\mathbf{v}$ is the velocity of the fluid considering the mesh as the rest frame. The following is the momentum equation, the Cauchy momentum equation with the magnetic force term:

\begin{equation}
    \rho \left(\frac{\partial \mathbf{v}}{\partial t} + \mathbf{v} \cdot \nabla \mathbf{v}\right) = - \nabla P + \frac{1}{\mu_0} \left( \nabla \times \mathbf{B} \right) \times \mathbf{B} + \nabla \cdot \mathbf{T},
\end{equation}

\noindent where $P$ is the pressure, $\mathbf{B}$ is the magnetic field, and $\mu_0$ is the magnetic permeability of free space. The stress tensor, $\mathbf{T}$, represents how the momentum is transported due to fluid viscosity. The magnetic field evolution is given by the induction equation:

\begin{equation}
    \frac{\partial \mathbf{B}}{\partial t} = \nabla \times \left(\mathbf{v} \times \mathbf{B} - \eta \nabla \times \mathbf{B}\right). 
\end{equation}

\noindent The last equation is the energy conservation equation:

\begin{equation}
    \frac{\partial e}{\partial t} + \nabla \cdot (e\mathbf{v})= - P \nabla \cdot \mathbf{v}
\end{equation}

\noindent where $e$ is the volumic internal energy. 

The initial conditions for the 2D Orszag-Tang vortex are the following: 

\begin{equation}
    \mathbf{v}_0 = \left(v_x, v_y \right) = \left(-\sin(2\pi y),  \sin(2\pi x) \right),
\end{equation}
\vspace{-3mm}
\begin{equation}
    \mathbf{B}_0 = \left(B_x, B_y \right) =  \frac{1}{\sqrt{4 \pi}} \left(-\sin(2\pi y),  \sin(4\pi x) \right),
\end{equation}

\begin{equation}
    \rho = \frac{25}{36 \pi}, \: \: \: \: P = \frac{5}{12 \pi}.
\end{equation}

\begin{figure*}[htbp]
 \centering
    \includegraphics[width=\textwidth]{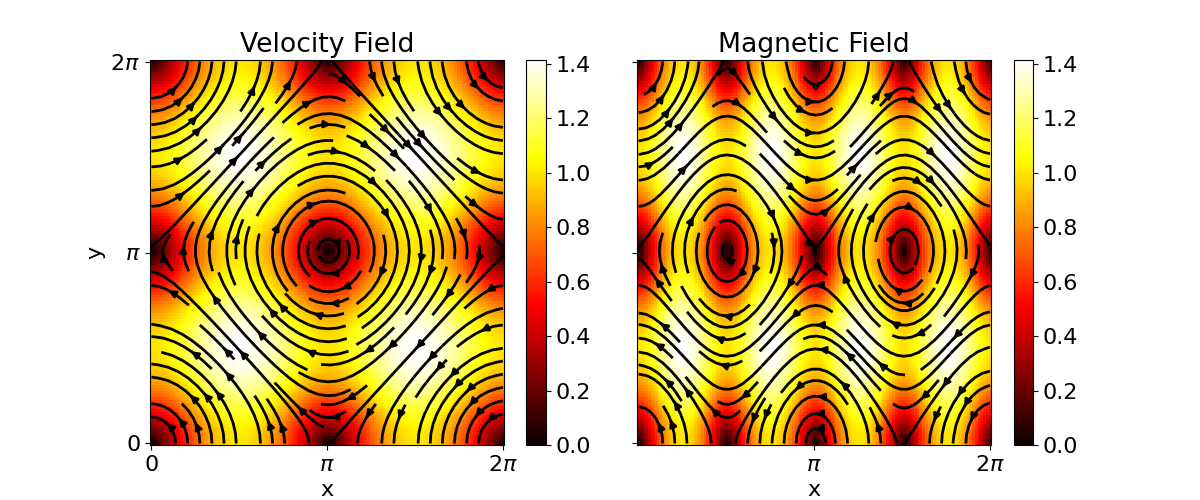}
    \caption{The initial velocity and magnetic field conditions in the Orszag-Tang vortex. The domain is given by $x \in [0, 2\pi]$ and $y \in [0, 2\pi]$. The colors are the intensity of each vector field. The plot shows that the test is initiated with vortices.}
    \label{initialcond}
\end{figure*}

\noindent With these parameters, the squared sound speed is given by $c_s^2 = \gamma P/\rho $, ensuring a normalized and consistent setup. The plasma has the following dimensionless quantities: the plasma beta $\beta$, the sonic Mach number $M_S$, and the Alfvénic Mach number $M_A$. These are computed as spatial averages over the entire domain, using the following expressions $ \beta =\frac{2\left\langle P\right\rangle}{ \left\langle B_x^2 + B_y^2\right\rangle}$, $M_S = \frac{\left\langle \sqrt{v_x^2 + v_y^2} \right\rangle}{\sqrt{\gamma \left\langle P \right\rangle / \left\langle \rho \right\rangle}}$ and $M_A = \frac{\left\langle \sqrt{v_x^2 + v_y^2} \right\rangle}{\sqrt{\left\langle B_x^2 + B_y^2 \right\rangle / \left\langle \rho \right\rangle}}$. For the simulations presented in this work, these quantities are evaluated as $\beta = 3.33$, $M_S = 0.95$, and $M_A = 0.78$. Figure \ref{initialcond} presents the initial magnetic and velocity fields within a domain of $x \in [0, 2\pi]$ and $y \in [0, 2\pi]$. These conditions will produce a compressible turbulent system, which is a transition to supersonic 2D MHD turbulence. 

\section{Methods}

In this section, we discuss the methods employed in this work. We begin by describing the numerical code utilized to evolve the Orszag-Tang vortex from its initial conditions. Next, we provide an overview of the FNO architecture. Finally, we detail the learning procedure implemented in the study.

\subsection{Numerical simulations}
\label{numericalsimulations}

To solve the Orszag-Tang vortex, we used the FARGO3D code \citep{Llambay2016}, which is designed to solve HD and MHD equations. Originally developed for protoplanetary disk simulations, FARGO3D was chosen for its GPU-oriented implementation, aligning with our ML model's GPU-based training and inference. Additionally, its astrophysical emphasis makes it a valuable tool for future applications in related fields.

FARGO3D employs the finite-difference method to discretize and solve continuous equations. This approach approximates solutions to PDEs by discretizing the domain into a grid and computing derivative approximations at these discrete points \citep{Stone1992, Zhou1993}. The finite-difference method often transforms PDEs into linear systems of equations that can be solved. To solve the conservation equations, FARGO3D uses an operator-splitting technique, wherein the PDE is decomposed into sub-problems that are solved sequentially.

The numerical
scheme uses substeps to update the
fields for each time-step. Initially, the source step is updated, followed by the transport step. For MHD simulations in FARGO3D, an additional substep is introduced to address the interaction between the flow and the magnetic field. This substep employs a constrained transport method \citep{evans1988} to ensure that the divergence of the magnetic field remains nearly zero. The constrained transport method uses the integral form of the induction equation to preserve magnetic flux conservation throughout the simulation.

To validate the robustness of FARGO3D, \cite{Llambay2016} investigated numerical tests, including several MHD-specific benchmarks. Among these tests is the Orszag-Tang vortex, which is a default setup in the FARGO3D code. The viscosity and diffusivity parameters are incorporated through the appropriate functions specified in the parameter file. These adjustments enable the study of how viscosity and diffusivity influence energy dissipation within the system. Since the original Orszag-Tang paper \citep{orszag1979} studying energy dissipation is a well-established test case for comparison.

\subsection{Neural operator}

Neural operators represent a class of ML models that generalizes neural networks (NNs) by learning an operator that maps between infinite-dimensional function spaces \citep{Kovachki2021}. Unlike traditional numerical solvers, which rely on the discretization of the solution space, neural operators learn an operator that maps functions in a continuous space. An operator $O$ maps one function space $F$ to another $G$,

\begin{equation}
    O: F \rightarrow G,
\end{equation}

\noindent  mapping an input function $f(x) \in F$ to an output function $g(x) \in G$, which is the solution of a PDE. Neural operators aim to approximate the mappings by constructing a parametric operator $O_\theta$ such as $O_\theta \approx O$. \cite{Kovachki2021} formalized the neural operator framework for learning operators. The operator $O$ is typically approximated by a deep neural network (DNN),  $O_\theta$, where $\theta$ represents the network parameters. The goal is to find the parameters $\theta^*$ that minimize the error between the predicted output function $g_\theta(x) =  O_\theta(f(x))$ and the true solution $g(x)$ given by

\begin{equation}
\theta^* = \text{arg min}_{\theta}  \; \mathcal{L}(g_\theta(x), g(x)),
\end{equation}

\noindent where $\mathcal{L}$ is the loss function. The network is trained on pairs of input-output functions,  $\{f_j, g_j\}^N_{j = 1}$, where each pair corresponds to a solution of the PDE for different initial or boundary conditions. Practical implementation involves their point-wise discretization since data in $\{f_j, g_j\}^N_{j = 1}$ represent functions.

The framework introduced by \cite{Kovachki2021} consists of lifting, iterative kernel integration, and projection. The architecture can be summarized as follows:

\begin{itemize}
\item \textbf{Lifting}:  The lifting step maps the input function $f(x) \in F$, defined in a low-dimensional space, into a higher-dimensional feature space for enhanced representation. It enables the model to capture complex dependencies and features from the input function. This transformation is achieved using a learned function, usually a neural network or a linear map
 
    \begin{equation}
     v_0(x) = \mathcal{Q}(f(x)), \;\;\;\;\; v_0(x) \in \mathbb{R}^{d_{v_0}}
    \end{equation}

\noindent where $\mathcal{Q}$ is the lifting operator, and $d_{v_0}$ is the dimensionality of the lifted feature space.

\item \textbf{Iterative Kernel Integration}: It is the mapping between the hidden representations. This step maps each hidden representation to the subsequent one. At each iteration $t$, the feature representation $v_t(x)$ is updated as

   \begin{equation}
        v_{t+1}(x) = \sigma_{t+1} \left[ W_t v_t (\Pi_t(x)) + (\mathcal{K}_t(v_t))(x) + b_t (x) \right]\;\;\; \forall x \in D_{t+1}.
    \end{equation}

\noindent This mapping involves a combination of a linear transformation, an integral kernel operator, and a bias. The linear transformation is a matrice multiplication, where the matrice is denoted as $W_t$. The integral kernel operator, $\mathcal{K}_t$, is a mapping given by

        \begin{equation}
    \label{integraloperator}
        (\mathcal{K}_t(v_t))(x) = \int_{D_t} \kappa^{(t)}(x,y) v_t (y) d\nu_t (y) \;\;\;\;\;\; \forall x \in D_{t+1},
    \end{equation}

where $\nu_t$ is a measure on $D_t$. The integral kernel operator is similar to the weight matrix in a NN since they both map from one function space to another. $\Pi_h$ is usually the identity, and $\sigma$ is the activation function. The bias, $b_h$, are also functions.

\item \textbf{Projection}: It maps the last hidden representation, $v_T$, to the output, $g$. In the code, the projection is executed via a linear transformation, similar to the process in the lifting phase,

  \begin{equation}
     g(x) = \mathcal{P}(v_T(x)), \;\;\;\;\; g(x) \in G
    \end{equation}

\noindent where $\mathcal{P}$ is the projection operator.

The equation summarizes the overall structure described by these three steps:

\begin{equation}
    \mathcal{G}_{\theta} := \mathcal{Q} \;\circ\; \sigma_H \left(W_{T-1} + \mathcal{K}_{T-1} + b_{T-1} \right)\;\circ\; ... \;\circ\; \sigma_1 \left(W_{0} + \mathcal{K}_{0} + b_{0} \right)\;\circ\;\mathcal{P}.
\end{equation}

\end{itemize}  

The main motivation behind neural operators is their capability to solve PDEs by learning mappings between infinite-dimensional function spaces \citep{Kovachki2021}.  This approach generalizes NNs to handle functional inputs and outputs since they are independent of the discretization of the data. Neural operators are general frameworks, as demonstrated by their capacity to generalize architectures like state-of-the-art Transformers \citep{Vaswani2017}, which are discussed to be a special case of neural operator in \cite{Kovachki2021}.

% Motivation for Fourier
Despite their potential, the integral kernel operator can introduce computational limitations, making it time-consuming and less competitive compared to conventional NNs. One strategy to address this challenge is to parameterize the integral kernel directly in the Fourier space, leading to the development of Fourier Neural Operators (FNOs) \citep{li2021}. Although alternative configurations of the integral kernel operator exist, this work focuses specifically on the Fourier-based formulation used in FNOs, which balances computational efficiency with modeling power.

\subsection{Fourier neural operator}

\begin{figure*}[htbp]
 \centering
    \includegraphics[width=\textwidth]{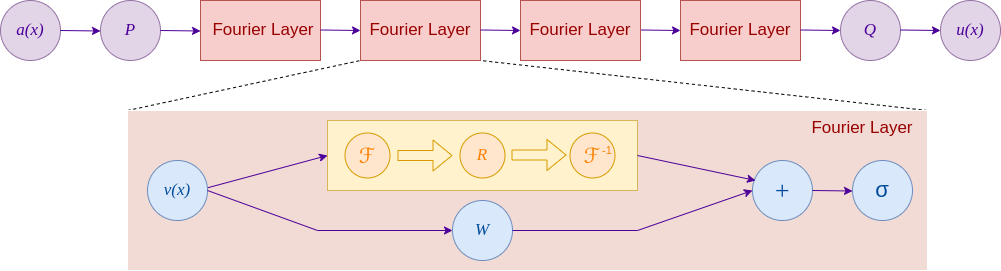}
    \caption{Scheme of a Fourier Neural Operator with 4 Fourier layers. $a(x)$ represents the data entering the lifting ($P$) part. The Fourier layer comprises two concurrent paths: the first is the path where a Fourier transform is applied, a matrix $R$ truncates the modes, and the inverse Fourier transform is applied. The second path is a convolution layer. Both results are summed, and an activation is applied. After four layers, the result goes to a projection ($Q$) part, and the outcome, $u(x)$, is finally computed. Inspired by \cite{li2021}.}
    \label{diagram}
\end{figure*}

FNOs is similar to neural operators, as both aim to learn mappings between infinite-dimensional function spaces, the difference being
the integral. The FNO architecture consists of key components: lifting, iterative kernel integration, and projection. Instead of using a standard integral kernel, the FNO implements the operator as a convolution, realized through a linear transformation in the Fourier domain. Figure \ref{diagram} outlines the FNO architecture, comprising the lifting operation $\mathcal{P}$, the projection $\mathcal{Q}$, and a sequence of four Fourier layers.

In an FNO, the integral operator from Equation \ref{integraloperator} is replaced by a convolution operator defined in Fourier space. For a given function $f(x)$, the Fourier transform $\mathcal{F}$ and its inverse $\mathcal{F}^{-1}$ are

\begin{equation}
    \mathcal{F}(k) = \frac{1}{(2\pi)^D} \int_D f(x) e^{-2i\pi\langle x,k\rangle} dx,
\end{equation}

\begin{equation}
    \mathcal{F}^{-1}(x) =  \frac{1}{(2\pi)^D} \int_D f(k) e^{2i\pi\langle x,k\rangle} dk.
\end{equation}

\noindent We can establish the Fourier integral operator as:

\begin{equation}
    (\mathcal{K}_h (v_h)) (x) = \mathcal{F}^{-1} (R_h \cdot (\mathcal{F}v_h)) (x),
\end{equation}

\noindent where $R$ is the transformation operator. 

The transformation operator $R$ in the Fourier space is represented as a tensor and it is similar to a weight matrix. In summary, the discrete Fourier transform is

\begin{equation}
    (R \cdot \mathcal{F}(v_h))_{k,l} = \sum_{j} R_{k, j, l} (\mathcal{F}(v_t))_{k, j}.
\end{equation}

The algorithm built to process the FNO follows these steps:

\begin{enumerate} 
    \item The data enters the algorithm and undergoes lifting into a higher dimension achieved through a linear transformation, yielding $v$. 

    \item Following the lifting stage, $v$  
 will go through two concurrent paths: (1) A Fourier transform is applied to $v$, followed by multiplication with an array $R$, and the modes will be truncated. An inverse Fourier transform is applied. (2) $v$ goes through a convolution. 

    \item The outcomes from both (1) and (2) paths are pointwise summed.

    \item An activation function is applied to the resulting sum.

    \item The result undergoes projection via another linear transformation, producing the output $g$.
    
\end{enumerate}

\noindent Note that the steps $2-4$ repeat depending on how many layers are defined in the algorithm. We show a FNO architecture in Figure \ref{diagram}.

\subsection{Learning procedure}

For the learning process, we utilized the PyTorch framework. In the original study by \cite{li2021}, experiments with the Navier-Stokes equation on a $64 \times 64$ grid demonstrated that the FNO captured the system's overall dynamics even with fewer Fourier modes. Building on this, our work employed a $128 \times 128$ spatial grid and found that the FNO performed optimally when truncated at 64 Fourier modes. The truncation of Fourier modes gives $-32 \leq  k_x, k_y \leq 32$. We used five Fourier modes in the temporal domain with ten timesteps. Our architecture included five Fourier layers, each with a width of $30$, and convolutional layers of the same size. We used the Adam optimizer \citep{kingma2017} with a learning rate of 0.001 for optimization. The loss function combined a modified L2 loss, as proposed by \cite{li2021}, with the mean absolute error (MAE). The training was conducted across various GPUs—NVIDIA RTX A5000, NVIDIA Quadro P6000, and NVIDIA Quadro GP100. Despite differences in hardware, the learning procedure consistently converged to similar results, underscoring the robustness of the model and framework.

\subsection{Evaluation}

The evaluation metrics used during the training and validation phases have been discussed earlier. Here, we focus on the metric employed during testing to quantify model's performance. For this purpose, we use the standard Mean Squared Error (MSE), defined as

\begin{equation}
\text{MSE} = \frac{1}{N} \sum_{i=1}^{N} (Y_i - \hat{Y}_i)^2
\end{equation}
 
\noindent where $Y_i$ is the target value and $\hat{Y}_i$ is the prediction. In addition to the MSE, we use the Structural Similarity Index (SSIM) as another evaluation metric. While MSE measures pixel-wise differences, SSIM evaluates structural similarity by considering luminance, contrast, and texture between the predicted and target fields. The SSIM is defined as:

\begin{equation}
    \text{SSIM}(x,y) = \frac{(2\mu_x\mu_y + c_1)(2\sigma_{xy} + c_2)}{(\mu_x^2 + \mu_y^2 + c_1)(\sigma_{x}^2 + \sigma_{y}^2 + c_2)}
\end{equation}

\noindent where $\mu_x$ and $\mu_y$ are the mean values of images, $\sigma_{x}^2$ and $\sigma_{y}^2$ are the variances, $\sigma_{xy}$ is the covariance, and $c_1$ and $c_2$ are small constants to prevent numerical instability. A higher SSIM value indicates that the prediction maintains the structural integrity of the target field. Another method for evaluating the results is through the power spectrum. This physical quantity indicates how energy is distributed across various scales within the flow. The power spectrum is defined as

\begin{equation}
 P(k) = \langle |\hat{\mathbf{f}}(\mathbf{k})|^2 \rangle 
\end{equation}

\noindent where $\hat{\mathbf{f}}(\mathbf{k})$ is the Fourier transform of the velocity or magnetic field

\begin{equation}
  \hat{\mathbf{f}}(\mathbf{k}) = \mathcal{F}(\mathbf{f}(\mathbf{r})) = \frac{1}{(2\pi)^2}\int_{\mathbb{R}^2} \mathbf{f}(\mathbf{r}) e^{-i\mathbf{k} \cdot \mathbf{r}} d\mathbf{r},
\end{equation}

\noindent and the brackets mean a directional integration in the $\mathbf{k}$ space $\sum_{| \mathbf{k} | = k}$. The power spectrum offers valuable insights into the model's performance across different scales, from large scales (low modes) to small scales (high modes). The Fourier-based nature of FNOs makes the power spectrum particularly well-suited for evaluating their predictions, as it aligns naturally with the spectral domain in which FNOs operate.

\section{Data}
\label{data}

In this section, we dicuss the dataset used for training and evaluating our model. First, we describe the data generation process using the FARGO3D code \citep{Llambay2016}. Next, we detail the data preparation steps required to format the data for input into the model. A full list of the simulations conducted is provided in Appendix A.

\subsection{Data generation}

The FARGO3D code implements the Orszag-Tang experiment as detailed in $\S$\ref{numericalsimulations}, with the initial conditions also described in that section. The original Orszag-Tang study \citep{orszag1979} focused on investigating how viscosity ($\nu$) and diffusivity ($\mu$) affect energy dissipation within the system. The energy dissipation rate is expressed as
\begin{equation}
\epsilon(t) = - \frac{dE}{dt} = \epsilon_K(t) + \epsilon_M(t)
\end{equation}
where 
\begin{align}
& \epsilon_K(t)=(\nu/2\pi )\int_0^{2\pi} \int_0^{2\pi} w^2 dxdy, \\
& \epsilon_M(t)=(\mu/2\pi )\int_0^{2\pi} \int_0^{2\pi} j^2 dxdy
\end{align}
are the kinetic and ohmic dissipation terms, respectively. The vorticity is given by  $\mathbf{w} = \nabla \times \mathbf{u}$, and the current
density is defined by $\mathbf{j} = \nabla \times \mathbf{B}$. 

Increased viscosity and diffusivity amplify energy dissipation by enhancing internal friction and transforming magnetic energy into heat, directly influencing turbulence and flow dynamics \cite{orszag1979}. To evaluate the FNO's capacity to capture these effects, we train the model across a range of viscosity and diffusivity values.

\begin{figure*}[htbp]
 \centering
\includegraphics[width=\textwidth]{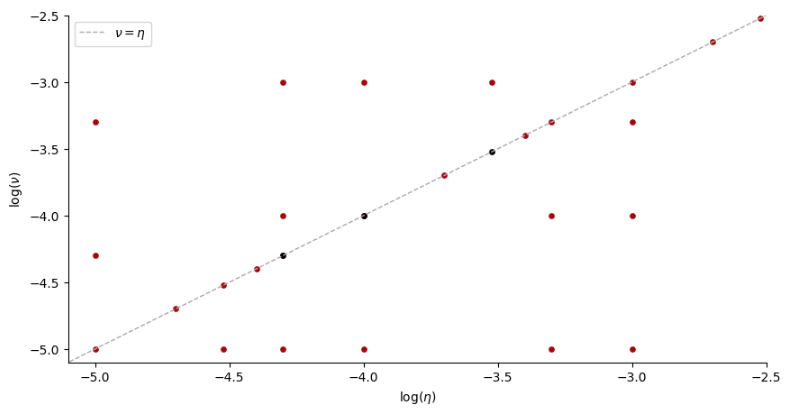}
    \caption{Plot showing all the combinations of $\nu$ and $\mu$ simulated. The red dots show the test set while the black dots show the training. The training set is divided between $80\%$ train and $20\%$ validation. The validation set is chosen randomly each training. 
}
    \label{numuappendix}
\end{figure*}

\begin{figure*}[htbp]
 \centering
    \includegraphics[width=\textwidth]{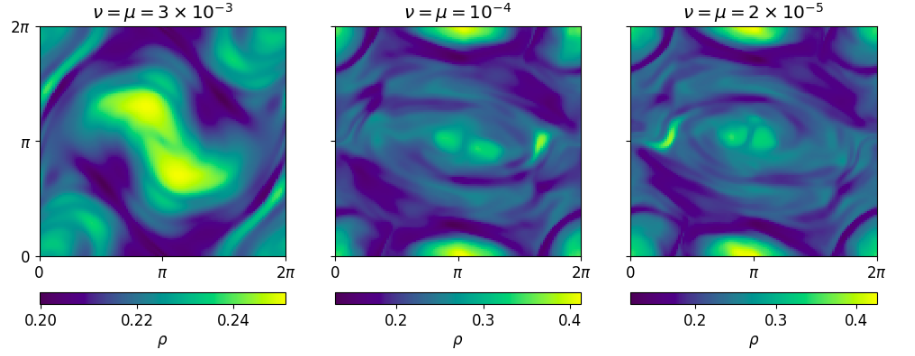}
    \caption{Panels show the density fields after 200 timesteps and how the viscosity and diffusivity impact the system's dynamic behavior. The values for viscosity nd diffusivity are shown at the top of each map.
}
    \label{training_data}
\end{figure*}

We simulated viscosity ranges, $\nu$, and diffusivity ranges, $\mu$, arbitrarily chosen in the interval $[10^{-5}, 5 \times 10^{-2}]$. Figure \ref{numuappendix} displays the preference for $\nu=\mu$ in this work is influenced by the original Orszag-Tang study. In total, we have 50 simulations with 1000 timesteps each. Figure \ref{training_data} shows that the timestep, t = 200, for the systems $\nu = \mu = 3 \times 10^{-3}$, $\nu = \mu = 10^{-4}$, and $\nu = \mu = 2 \times 10^{-5}$. While all simulations contribute to training and validation, simulations with $\nu = \mu = 5 \times 10^{-5}$ and $\nu = \mu = 3 \times 10^{-4}$  are reserved for evaluation purposes.

\subsection{Data preparation}

Data preparation defines how information is structured and supplied to the model. For the 3D-FNO architecture, the input data must conform to a five-dimensional block format, $(N, W, H, T_{\rm in}, C)$, where  $N$ represents the number of batches, $(W, H)$ are the spatial dimensions, $T_{\rm in}$ represents the temporal dimension, and $C$ (channels) encapsulates spatial, temporal, and physical system information. In this study, $C$ has a size of 7, including 5 input frames and 2 physical parameters ($\nu$, $\mu$). The output data, meanwhile, is formatted as a four-dimensional block, $(N, W, H, T_{\rm out})$, where the spatial dimensions are identical to the input, but $T_{\rm out}$ includes 10 frames.

\begin{figure*}[htbp]
 \centering
 \includegraphics[width=0.8\textwidth]{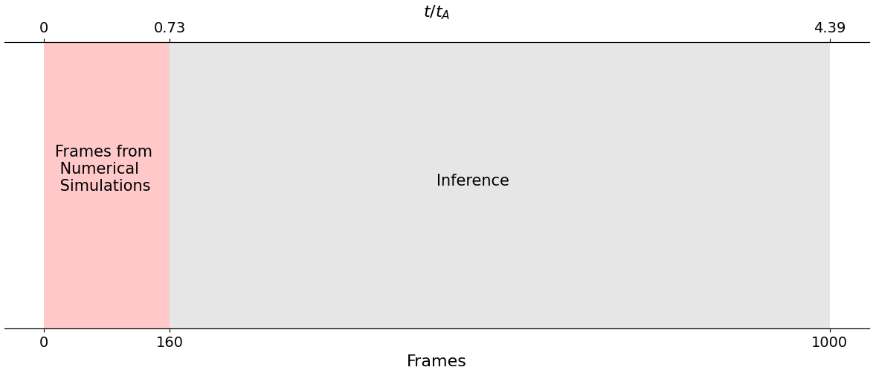}
    \caption{The goal is to have a model that is trained using the initial 160 frames $\Delta t = 0.73 \:t_A$) and is capable of predicting the sequence from frame 160 to 1000 $(0.73$ to $4.39 \:t_A)$. All systems are trained with the same configuration. }
    \label{temporal}
\end{figure*}

Through experimentation, we identified an optimal configuration balancing performance and speed-up. The selected input block size was $(16, 128, 128, 10, 7)$  and the output block size was $(16, 128, 128, 10)$. Namely, $N$ is set to $16$, the simulation grid spans $128 \times 128$, input data consists of 5 frames spaced $t = 1.0$ code units apart, and output data consists of 10 frames with intervals of $t = 4.0$ code units. The time step between consecutive frames is $\Delta t = 0.05$. In practice, this means the model trains on the interval $[0, 0.73]\:t_A$ using 160 frames and predicts the interval $[0.73, 4.39]\:t_A$ spanning frames 160 to 1000. Figure \ref{temporal} illustrates how the model predicts the system evolution based on this configuration.

Normalization is an important step in data preparation, ensuring that all features are scaled to a consistent range. This process improves performance and stability by ensuring that features contribute equally to the training process \citep{Singh2020}. Normalization also helps the network focus on components that carry significant information \citep{Huang2020}. For this work, we adopted a modified version that scales values to the range $[-1, 1]$. The scaling process is expressed as:

\begin{equation}
    x' = a + b \frac{x - x_{\rm min}}{x_{\rm max} - x_{\rm min}},
\end{equation}

\noindent where $a$ is $-1$ and $b$ is $2$ in our approach. Here,  $x'$ represents the normalized feature, which could correspond to variables such as density, velocity field, or magnetic field.

\section{Results}
\label{results}

In this section, we present the results of the trained model. The results will be presented in the following order: instantaneous predictions, followed by the temporal evolution of both the velocity and magnetic fields. Finally, we will examine properties that involve both variables. The analysis employs a time unit called the Alfvén time ($t_A$), which represents the time required for an Alfv\'en wave to traverse a characteristic length scale ($L$) in a plasma. Alfv\'en waves, a type of wave in the MHD regime, which is a transverse non-dispersive wave that propagates with phase speed $\propto \cos \theta$, where $\theta$ is the angle
between the direction of the wave-propagation (wave-vector) and the local background magnetic field.  The Alfv\'en time is defined as $t_A = L/v_A$, where $v_A$ is the Alfv\'en velocity, given by the initial magnetic field strength ($B_0$) and the initial plasma density ($\rho_0$), $v_A = \sqrt{B_0^2/4 \pi \rho_0}$. In our study, one timestep corresponds to approximately $1 \;t_A \sim 0.0045$ in code units.

In this section, we present the results for $\nu = \eta = 5 \times 10^{-5}$, corresponding to a system characterized by low viscosity and resistivity. Additional results for different parameter sets and grid resolutions are provided in Appendix \ref{app:appendixB} to evaluate the model's robustness and generalization capability.

\subsection{Velocity Field}
\label{sec:velocityfield}

Each velocity component is considered, and we will discuss the results in terms of the velocity magnitude, given by $|\mathbf{u}| = \sqrt{u_x^2 + u_y^2}$, and the streamlines. The component results are available in Appendix \ref{app:appendixC}. The analysis of the velocity field will be presented from two perspectives: first, by directly comparing the system state at $t = 1\; t_A$, which is $\sim 200$ frames ahead, and second, by examining the temporal evolution of $\mathbf{u}$.

\subsubsection{Instantaneous comparison}

\begin{figure*}[htbp]
 \centering
 \includegraphics[width=\textwidth]{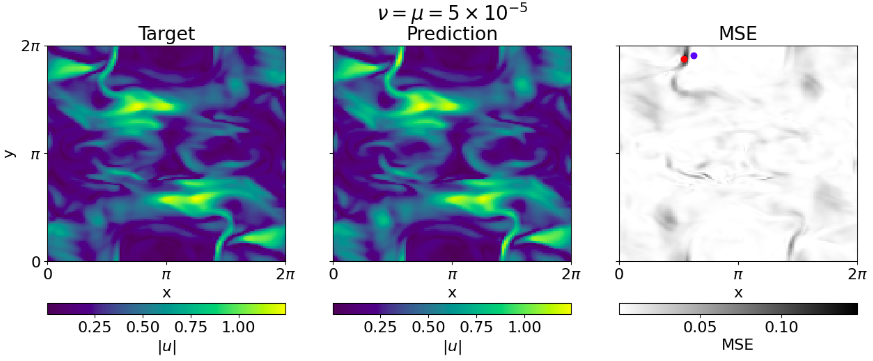}
    \caption{The figure compares $|\mathbf{u}|$ at $1\;t/t_A$. The first panel displays the target field, the second shows the model's prediction, and the third presents the MSE, where darker regions indicate higher errors. The blue and red dots mark the locations of the minimum and maximum errors, respectively. The discrepancies are concentrated in a region where a discontinuity is present, highlighting the challenge of capturing abrupt changes.}
    \label{results:u}
\end{figure*}

Figure \ref{results:u} compares $|\mathbf{u}|$ at $t = 1\; t_A$ between the target and the prediction. Visually, the target and prediction are nearly indistinguishable. The figure also includes the MSE map, highlighting the regions of lowest (blue dot) and highest (red dot) error. Both dots correspond to areas where a shock---an abrupt change in $|\mathbf{u}|$---is observed. These errors highlight a limitation of NNs, including FNOs, as they struggle to accurately capture sharp gradients and discontinuities, a challenge associated with the Gibbs phenomenon. Table \ref{results:velocity:tableu} summarizes some evaluation metrics for Figure \ref{results:u}, including the values for the blue (minimum error) and red dots (maximum error), as well as the mean and median errors.

\begin{table}[h]
\begin{center}
\begin{tabular}{|l|l|l|}
\hline
 & $|\mathbf{u}|$ &  $|\mathbf{B}|$ \\
\hline
Maximum Error & $1.46 \times 10^{-1}$ & $6.7 \times 10^{-2}$ \\ \hline
Minimum Error & $1.17 \times 10^{-9}$  & $3.6 \times 10^{-11}$ \\ \hline
Mean Error & $5.7 \times 10^{-3}$ & $1.21 \times 10^{-3}$ \\ \hline
Median & $2.59 \times 10^{-3}$ & $6.45 \times 10^{-4}$\\ \hline
SSIM Index & $0.92$ & $0.93$  \\ \hline
\end{tabular}
\caption{The table presents error metrics of $|\mathbf{u}|$ and $|\mathbf{B}|$ for the model's performance.} 
\label{results:velocity:tableu}
\end{center}
\end{table}

\subsubsection{Time Evolution}

\begin{figure*}[htbp]
 \centering
 \includegraphics[width=\textwidth]{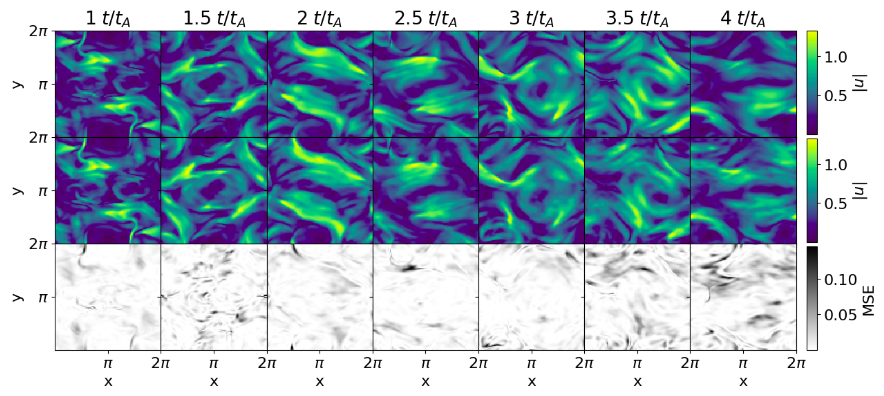}
    \caption{The evolution of the velocity field magnitude, $|\mathbf{u}|$, is shown across seven time steps. The first row represents the target values, the second row displays the model's predictions, and the third row illustrates the MSE. Darker areas in the MSE row correspond to regions with higher errors, occurring near areas of rapid change.}
    \label{results:uev}
\end{figure*}

Figure \ref{results:uev} illustrates the evolution of $|\mathbf{u}|$ across seven time steps. The first row displays the target magnitudes, the second row shows the predicted values, and the third row presents the MSE. The model successfully captures the overall structure and dynamic behavior of the field, though minor discrepancies arise in regions with rapid changes, where the MSE is higher. Towards the later time steps, darker regions indicative of higher errors are observed at smaller scales or high wavenumbers.

\begin{figure*}[htbp]
 \centering
 \includegraphics[width=\textwidth]{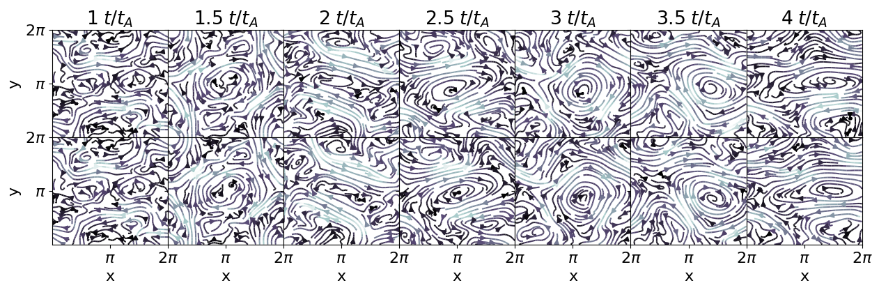}
    \caption{The evolution of the vectorial velocity field is presented, with the target shown in the first row and the prediction in the second. Both the target and prediction exhibit a consistent trend of vortex merging, developing a larger coherent vortex structure at later time steps.}
    \label{results:uvec}
\end{figure*}

Figure \ref{results:uvec} depicts the vectorial velocity field over the same time steps using streamlines. The first row represents the target field, while the second row shows the model’s predictions. Both the target and predicted fields exhibit the same trend of smaller vortices merging and evolving into larger vortex structures over time. This merging process highlights the model's capability to replicate turbulent dynamics and the formation of larger coherent structures.

\begin{figure*}[htbp]
 \centering
 \includegraphics[width=\textwidth]{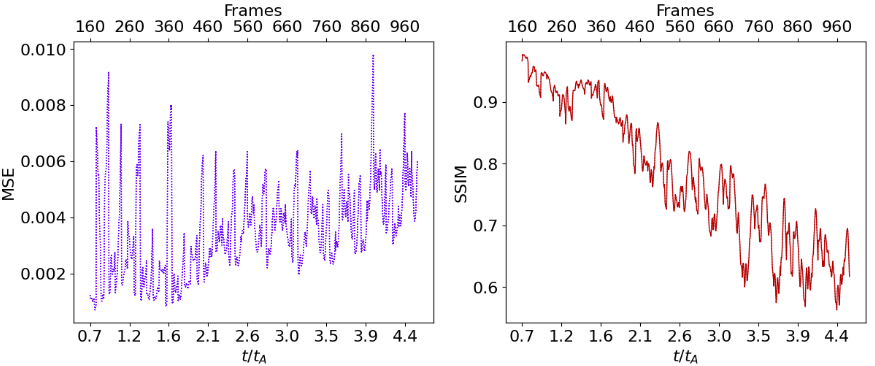}
    \caption{The MSE and SSIM plots show the model's predictive accuracy over time for $|\mathbf{u}|$. The MSE plot reveals a growing error as time advances, with peaks associated with small-scale regions. Similarly, the SSIM plot exhibits a decline in structural similarity over time, reflecting a decrease in the model's ability to preserve spatial patterns.}
    \label{results:verror}
\end{figure*}

In Figure \ref{results:verror}, we present the error metrics, including MSE and SSIM, over time. Both metrics indicate a clear trend: prediction accuracy diminishes as time progresses, with increasing discrepancies between the target and predicted values. Peaks in the MSE plot correspond to regions dominated by small-scale features. The peaks highlight the challenge of accurately capturing these finer structures. Similarly, the SSIM plot shows a gradual decrease over time. 

\subsection{Magnetic Field}

Similarly to the analysis done to $\mathbf{u}$, we will analyze the results in terms of the magnetic field magnitude, given by $|\mathbf{B}| = \sqrt{B_x^2 + B_y^2}$, and the magnetic field lines. 

\subsubsection{Instantaneous comparison}

\begin{figure*}[htbp]
\centering
 \includegraphics[width=\textwidth]{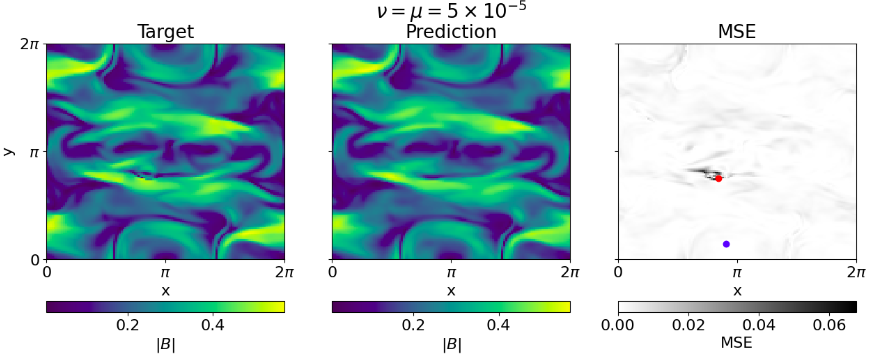}
    \caption{Comparison of the magnetic field magnitude $|\mathbf{B}|$ between target and prediction at $1\;t/t_A$. Convention as Figure  \ref{results:u}. The blue dot marks the location of the lowest error while the red dot indicates the highest error.}
    \label{results:b}
\end{figure*}

In Figure \ref{results:b}, we compare the predicted magnetic vector field to the target, highlighting that the highest errors occur in regions of discontinuity. In areas with abrupt field changes, the maximum error reaches approximately $6.7 \times 10^{-2}$, while smoother regions exhibit a much smaller error of $\sim 10^{-11}$. These results emphasize the model's difficulty in accurately capturing fine details in regions with rapid changes. Detailed error metrics can be found in Table \ref{results:velocity:tableu}.

\subsubsection{Time Evolution}

\begin{figure*}[htbp]
 \centering
 \includegraphics[width=\textwidth]{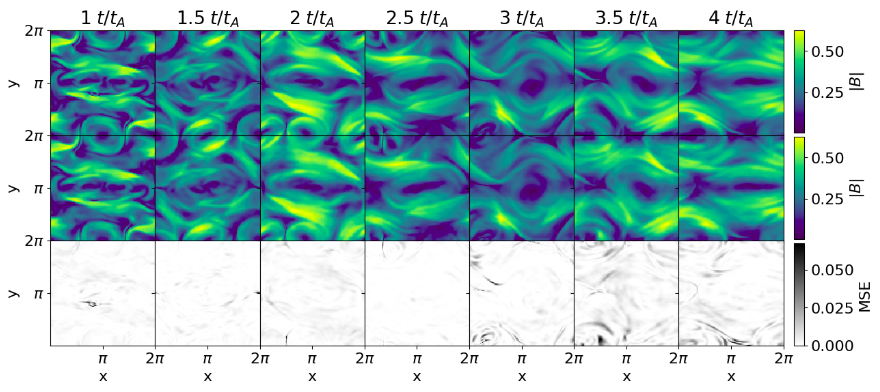}
    \caption{Comparison of $|\mathbf{B}|$ between the target and prediction across seven timesteps. Convention as Figure \ref{results:uev}. }
    \label{results:bev}
\end{figure*}

\begin{figure*}[htbp]
 \centering
 \includegraphics[width=\textwidth]{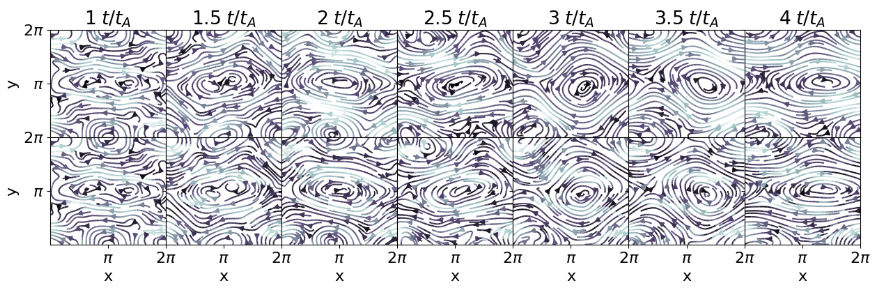}
    \caption{Comparison of the magnetic field vector $\mathbf{B}$ between the target and prediction over seven timesteps as Figure \ref{results:uvec}.}
    \label{results:bvec}
\end{figure*}

Figure \ref{results:bev} compares the target and predicted magnetic field magnitudes across seven timesteps, showcasing the model's ability to capture the field's evolution. In the initial stages, the target and predicted values align closely, with the MSE approaching zero. However, as time progresses, subtle discrepancies emerge, particularly in regions with higher gradients. Figure \ref{results:bvec} focuses on the magnetic lines, illustrating the directional behavior and structure of the field. Both the target and predicted fields reveal a similar evolution of magnetic lines.

\begin{figure*}[htbp]
 \centering
 \includegraphics[width=\textwidth]{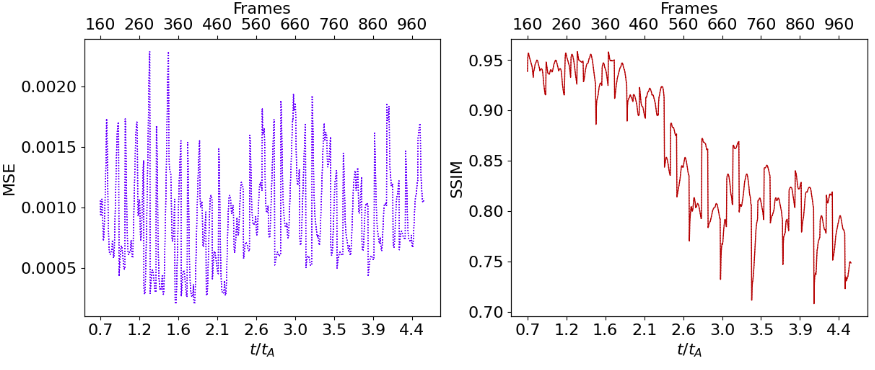}
    \caption{The plot displays the MSE and SSIM metrics for the magnetic field predictions compared to the target over time. The MSE presents peaks corresponding to specific moments, driven by dynamic changes in the field. The SSIM shows a decreasing trend, with variations peaking at certain points, suggesting fluctuations in the model's performance, potentially influenced by the structure of the training data.}
    \label{results:berror}
\end{figure*}

We present the MSE and SSIM metrics over time in Figure \ref{results:berror}. The first plot shows that the MSE gradually increases as the simulation progresses, with the peaks at specific intervals. Similarly, the second plot indicates a decline in SSIM over time, accompanied by peaks of variation in accuracy. These fluctuations in both MSE and SSIM suggest that the model's training methodology did not fully account for periodic boundary conditions along the temporal dimension. This observation highlights the need to refine the training process to better preserve periodicity over time.

\subsection{Energy Dissipation}

The original study by \citep{orszag1979} investigates the influence of $\nu$ and $\eta$ on fluid dynamics.  Higher values of $\nu$ amplify dissipation, as friction converts kinetic energy into heat \citep{Richardson2007}. The energy is transferred from larger to smaller scales via turbulent eddies \citep{Goldreich1995}. However, it is important to note that in 2D simulations, the energy spectrum differs from that in 3D simulations. The energy spectrum will be discussed in detail in the next section. Similarly, $\eta$ determines the diffusion or spread of the magnetic field within the medium. In MHD, higher magnetic diffusivity is associated with increased magnetic energy dissipation, leading to the conversion of magnetic energy into heat. 

%One way to evaluate this impact is by analyzing the energy dissipation, defined as:
%\begin{equation}
%\epsilon = \frac{\nu}{2\pi} \int_0^{2\pi} \int_0^{2\pi} w^2 dxdy  +  \frac{\eta}{2\pi} \int_0^{2\pi} \int_0^{2\pi} j^2 dxdy,
%\label{kinematicdiss}
%\end{equation}

%\noindent where $w = \nabla \times \mathbf{u}$ is the vorticity and $j = \nabla \times \mathbf{B}$ is the current. Since $\mathbf{u}$ and $\mathbf{B}$ are two-dimensional vectors, $w$ and $j$ are scalars. As shown in Equation \ref{kinematicdiss}, $\nu$ and $\eta$ directly influence the magnitude of the dissipation.

\begin{figure*}[htbp]
 \centering
 \includegraphics[width=\textwidth]{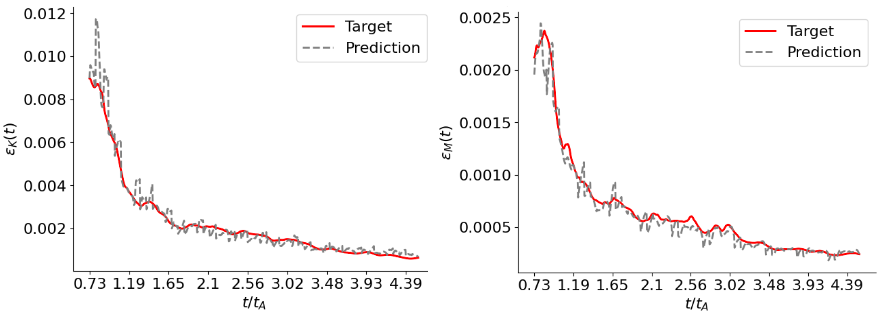}
    \caption{Kinetic and magnetic energy dissipation plots. The first plot compares the target and predicted kinetic energy dissipation, showing similar trends. The second plot presents magnetic energy dissipation, which initially peaks and then gradually decreases.}
    \label{results:energys}
\end{figure*}

Figure \ref{results:energys} presents the
time series for kinetic and magnetic energy dissipation. In the first plot, the kinetic energy dissipation for both the target and the prediction follows a similar trajectory, starting with a peak and gradually asymptoting toward zero. This alignment indicates that the prediction closely follows the results reported by \cite{orszag1979}. However, the prediction exhibits slight fluctuations compared to the target, which can be attributed to errors at smaller scales (higher wavenumbers). Since kinetic energy dissipation is associated with these small scales, these fluctuations highlight the model's limitations in fully resolving finer details. Nonetheless, the model effectively replicates the overall dissipation trend. The right-hand plot in Figure \ref{results:energys} illustrates the dissipation of magnetic energy over time. Similar to kinetic energy dissipation, the magnetic energy dissipation starts with a peak and decreases asymptotically. This behavior reflects an initial burst of magnetic energy dissipation, after which the dissipation rate slows as the system stabilizes. These plots indicate the model's ability to capture the overall dissipation dynamics characteristic of the Orszag-Tang vortex.

\begin{figure}[h]
 \centering
 \includegraphics[width=\columnwidth]{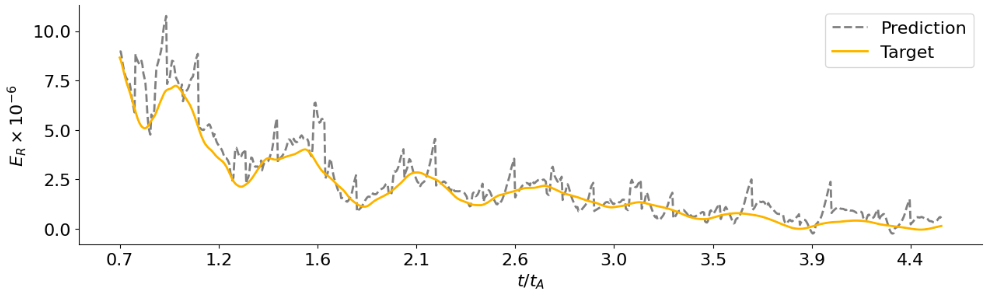}
    \caption{The plot compares the residual energy predicted by the model (gray dotted line) to the target residual energy (orange line).}
    \label{results:energy}
\end{figure}

In Figure \ref{results:energy}, we analyze the residual energy. Residual energy in MHD fluids is the difference between kinetic and magnetic energy, defined as $E_R = E_k - E_m$. It gives insights into the energy balance between fluid motion and magnetic fields, influencing turbulence dynamics and cascade processes. We compare the predicted residual energy with the target residual energy. Although spikes are observed in the model's residual energy, due to the challenge to model finer structures, both the model and target display the same overall trend. These results indicate that the model captures the dominant features of the energy decay process, even with some deviations at smaller scales.

\subsection{Spectral Analysis}

Spatial spectral analysis is an important tool in fluid dynamics. It reveals, for example, how energy is distributed across various spatial scales. Since our model uses the Fourier domain as its learning space, spectral analysis is particularly useful for comparing target and predicted results. The 1D power spectrum is
defined as

\begin{equation}
P(k) = \sum_{|\mathbf{k}| = k} |\hat{f}(\mathbf{k})|^2
\end{equation}

\noindent where $\hat{f}(\mathbf{k})$ is the Fourier transform of some variable (for example, $\mathbf{u}$ or $\mathbf{B}$), $\mathbf{k}$ is the wavevector and $k = |\mathbf{k}| = \sqrt{k_x^2 + k_y^2}$ is the corresponding wave number. The power spectrum of the vector field can be interpreted in terms of the energy distribution across scales. The 1D power spectrum is a function of $k$, where higher $k$ values correspond to smaller spatial scales.

\begin{figure*}[htbp]
 \centering
 \includegraphics[width=\textwidth]{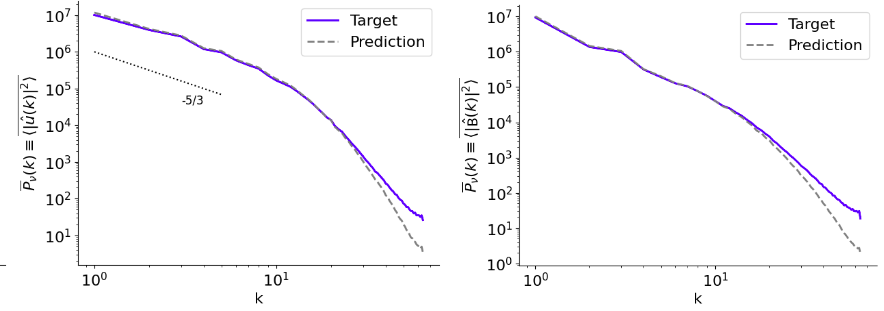}
    \caption{The left plot shows the mean power spectrum for the velocity field. At low wavenumber, the prediction aligns well with the target, while at higher wavenumbers (smaller scales), the prediction diverges. The dotted line represents the Kolmogorov $-5/3$ law, which describes the expected energy scaling in the inertial range of turbulence.  The same is observed in the right plot that shows the mean power spectrum for the magnetic field, where the model captures the magnetic energy distribution at low wavenumbers well but fails to replicate features at high wavenumber. This discrepancy can be attributed to the truncation of Fourier modes in the FNO architecture, which limits the model’s ability to capture fine-scale features. Since high wavenumbers correspond to smaller spatial structures, their exclusion from the spectral representation constrains the predictive resolution at those scales.}
    \label{results:spectrum}
\end{figure*}

The left plot in Figure \ref{results:spectrum} examines the mean power spectrum for the velocity field. The prediction aligns with the target at low wavenumbers (large scales), demonstrating accurate reproduction of large-scale structures. However, as wavenumber increases (smaller scales), the prediction diverges, showing the model's limitations
in reproduce fine details, at the smaller scales represented by the grid, already in the viscous/ohmic dissipative scale. Similarly, the second plot presents the mean power spectrum for the magnetic field. While the model effectively captures the magnetic energy distribution at low wavenumber, it deviates from the finer-scale structures. Note that during the training, there is truncation of the Fourier modes representing the smallest scales (higher wavenumber). Therefore, the prediction is not expected to accurately capture the small-scale structures. This discrepancy underscores the challenges of accurately modeling small-scale phenomena in velocity and magnetic fields. Another characteristic is the $-5/3$ law observed in the dotted line. The scaling -5/3 law, derived from Kolmogorov’s turbulence theory \citep{kolmogorov1941}, describes how kinetic energy cascades in an inertial range. However, this relation is not perfectly maintained because 2D MHD deviates from the $-5/3$ slope due to the different nature of turbulence in two dimensions. 

\begin{figure*}[htbp]
 \centering
 \includegraphics[width=\textwidth]{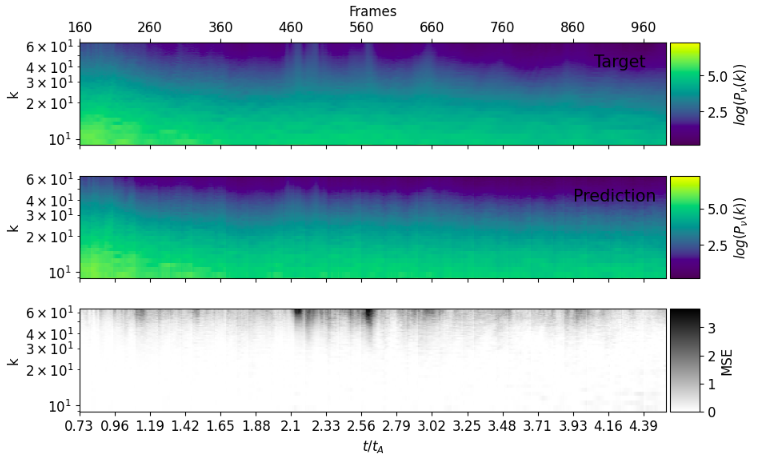}
    \caption{The velocity field spectrogram shows the model's performance over time. The lower-frequency components demonstrate the model's strength in representing larger structures, while its performance diminishes in capturing small-scale variations. This limitation is likely due to the truncation of Fourier modes inherent in the FNO architecture, which restricts the model’s capacity to represent fine-scale dynamics.}
    \label{results:spectrogram_v}
\end{figure*}

The spectrogram in Figure \ref{results:spectrogram_v} illustrates how the model captures the dynamics of different scales over time. Large-scale structures are effectively reproduced, as indicated by the low MSE in the lower-frequency components of the spectrogram. Over time, the energy density in small scales diminishes, and the model follows this trend accurately. While the model maintains consistent performance across all time steps, its limitations are most evident in capturing small-scale features.

\begin{figure*}[htbp]
 \centering
 \includegraphics[width=\textwidth]{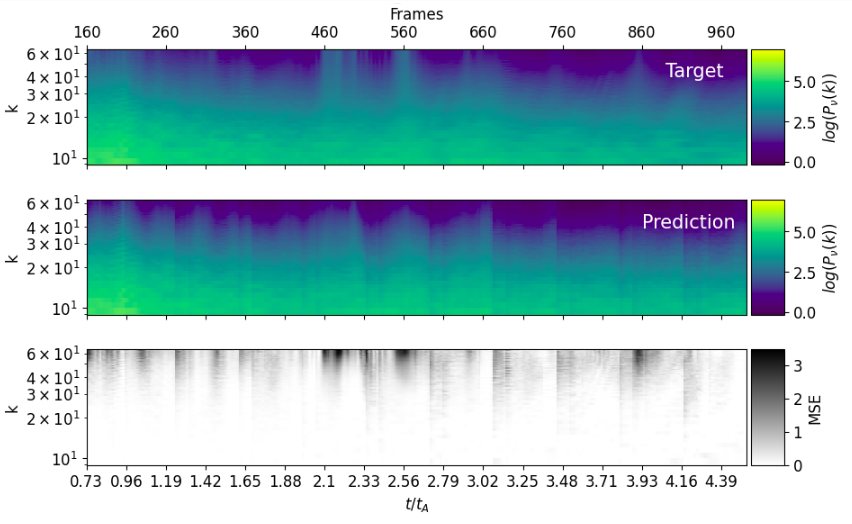}
    \caption{Magnetic field spectrogram comparison illustrating the target and predicted magnetic energy distribution over time.}
    \label{results:spectrogram_b}
\end{figure*}

The magnetic spectrogram in Figure \ref{results:spectrogram_b} shows how the model captures the energy distribution in the magnetic field over time. Both the target and prediction demonstrate a progressive shift of magnetic energy toward larger scales, reflecting the merging of smaller vortices into larger structures as the simulation advances. This vortex merging process, characteristic of the Orszag-Tang vortex evolution, results in energy becoming increasingly concentrated at lower wavenumbers. The figure indicates that the model successfully follows this pattern, suggesting its effectiveness in accurately predicting the energy distribution.

\subsection{Analysis of Density and Cross-Helicity}

\begin{figure*}[htbp]
 \centering
 \includegraphics[width=\textwidth]{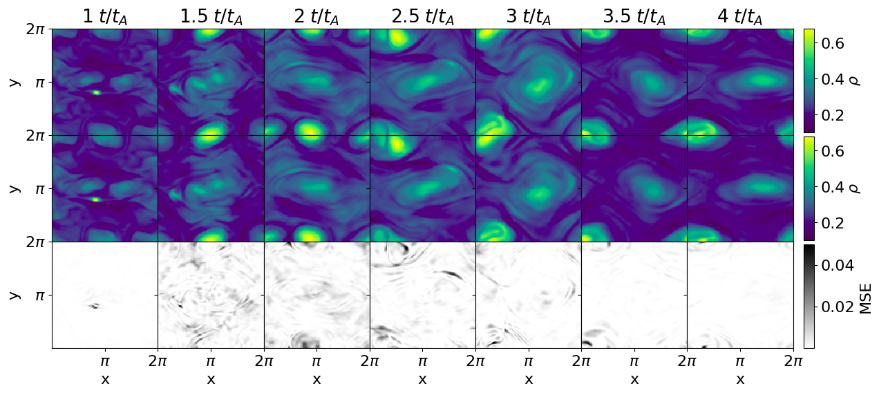}
    \caption{Evolution of density across seven timesteps, capturing the structural changes in the system over time. The third row displays the MSE between the target and the prediction for each timestep. Darker regions in the MSE plots highlight areas with higher prediction errors.}
    \label{results:rho}
\end{figure*}

Here, we will explore additional analysis. Starting with the density evolution shown in Figure \ref{results:rho}, we compare the target and predicted densities across various timesteps. The accompanying MSE highlights regions of deviation, revealing that discrepancies are predominantly concentrated in areas with small-scale structures and discontinuities. This suggests that the model exhibits heightened sensitivity to fine-scale variations.

\begin{figure*}[htbp]
 \centering
 \includegraphics[width=0.7\textwidth]{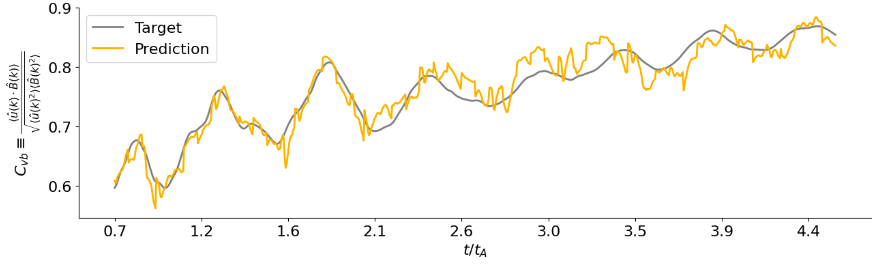}
    \caption{This figure compares the cross-helicity between the model's predictions and the target. Both show a similar trend, with the magnetic and velocity fields aligning more closely over time. This demonstrates the model's ability to capture the overall alignment trend.}
    \label{results:cross}
\end{figure*}

Figure \ref{results:cross} compares the cross-helicity between the model's predictions and the target. Cross helicity is a measure of the correlation between the velocity and magnetic fields, defined as $H_c = \langle \mathbf{u} \cdot \mathbf{B} \rangle$. It quantifies the degree of alignment between these fields, influencing turbulence dynamics and energy transfer. Both exhibit a similar trend, with the magnetic and velocity fields progressively aligning as the system evolves. This alignment highlights the model's capability to capture the interactions between the fields over time. However, discrepancies become apparent in the later stages of the simulation. Despite these differences, the model preserves the overall trend toward field alignment.

\subsection{Speed Up}

We evaluate the speed-up achieved with the FNO model using an NVIDIA GeForce RTX 3080 GPU. To ensure a fair comparison, both the FNO model and the numerical solver were run on the same hardware using the GPU-optimized fluid dynamics code \code{FARGO3D}. Model predictions were generated on the RTX 3080, while hyperparameter tuning was performed on a cluster with two NVIDIA RTX A5000 GPUs. The hyperparameter-tuned training consisting of 2,000 epochs, was executed on a high-performance cluster with NVIDIA Quadro P6000 and Quadro GP100 GPUs.

Our FNO model completes 800 frames in just 8 seconds, compared to 198 seconds required by the optimized numerical solver—achieving a speed-up of approximately $24.75 \times$. This substantial gain demonstrates the potential of FNO for faster MHD simulations, reducing computational costs while maintaining accuracy.

\subsection{Baseline Comparison}

Comparing different architectures for forecasting turbulent nonlinear environments is important for evaluating the strengths and limitations of FNO in capturing complex dynamics. In this study, we compare FNO with UNet, a convolutional encoder-decoder architecture widely used in fluid dynamics predictions. To ensure a fair comparison, both models were trained and tested on the same dataset with a resolution of $64 \times 64$, following the data preparation described in \ref{data}. The training was performed on an NVIDIA GeForce RTX 3080 GPU, and each model was trained until the loss function converged. The UNet implementation follows the architecture proposed by \cite{Duarte2022}. Our results show that FNO outperforms UNet in capturing both large-scale structures and finer details of the turbulent flow. While UNet is a powerful architecture for spatial feature extraction, it struggles with preserving temporal information. This suggests that despite its ability to learn complex spatial patterns, UNet may not be the best choice for highly dynamic, non-linear systems such as turbulence. 

\begin{figure}[htbp]
 \centering
 \includegraphics[width=\columnwidth]{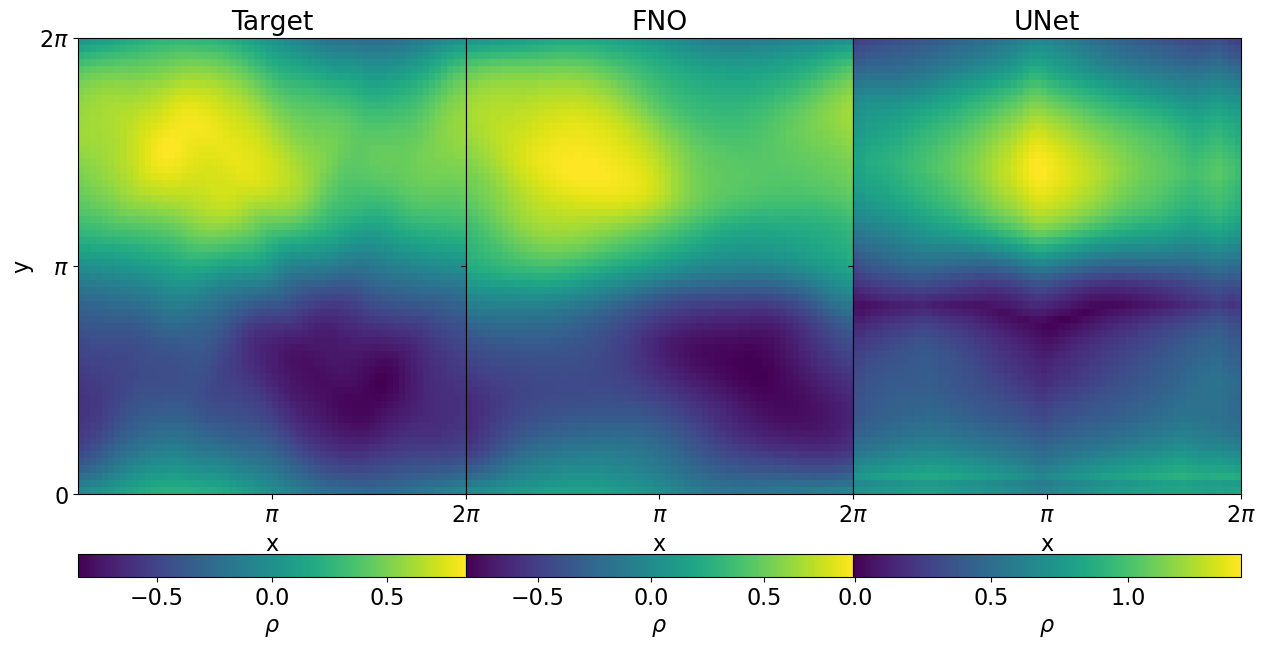}
    \caption{Comparison of the target solution with predictions from FNO and UNet at $1\;t/t_A$. The first plot shows the target, the second plot shows de FNO and the third plot shows the UNet. The FNO prediction is similar to the target capturing both large and small-scale structures. The UNet prediction appears smoothed.}
    \label{results:comparison}
\end{figure}

Figure \ref{results:comparison} presents three plots comparing the target solution with the predictions from FNO and UNet at $1\;t/t_A$. Visually, the FNO prediction matches the target, capturing both large-scale structures and fine-grained details of the turbulent flow. In contrast, the UNet output appears slightly different, suggesting that the model may have learned a smoothed representation, potentially averaging between frames despite the use of a gradient-based loss function. This behavior indicates that while UNet is useful at reconstructing spatial features, it may struggle with preserving the intricate temporal dynamics of highly non-linear turbulent flows.

\begin{figure*}[htbp]
 \centering
 \includegraphics[width=0.7\textwidth]{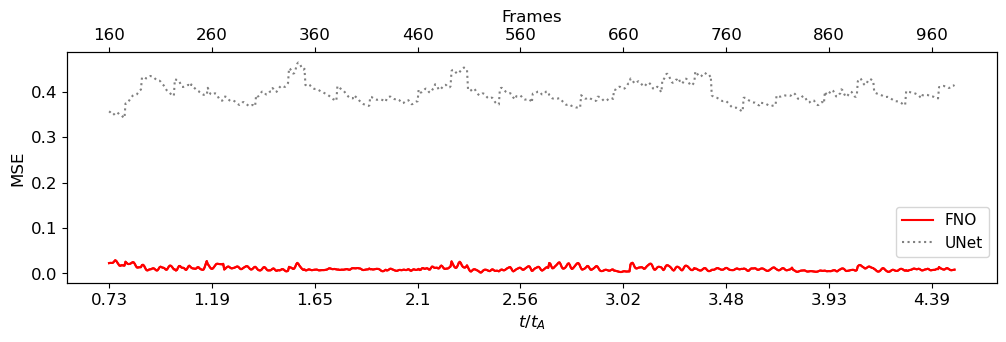}
    \caption{MSE as a function of time for each model with FNO (red) and UNet (gray). As expected, FNO exhibits the lowest error across all time steps. The UNet model presents higher error during all the timeline.}
    \label{results:comparison_mse}
\end{figure*}

Figure \ref{results:comparison_mse} presents the MSE as a function of time, $t$, for the FNO and UNet models. As expected, FNO achieves the lowest error across all time steps, highlighting its capability in capturing the complex dynamics of turbulence. The UNet model exhibits a higher error, confirming that it may be averaging the solution rather than preserving the structures of the flow. This trend aligns with the qualitative observations from the predictions and further reinforces the advantages of using FNO for forecasting non-linear turbulent environments.

\section{Discussion}

FARGO3D generates binary DAT files for each component, producing vectors of size $16,384$. These vectors were reshaped into 2D arrays and normalized using feature scaling. For 3D simulations with FNO, the input requires a five-dimensional block that integrates spatial, temporal, and physical information. Each input block includes temporal data from five frames with a timestep of $20$, while the output corresponds to ten temporal frames with a timestep of $80$. This block configuration is designed to optimize simulation acceleration while minimizing numerical errors. Although the model benefits from long-term forecasting capabilities, a trade-off exists in the form of reduced accuracy at the end of each block. This issue arises because the temporal dimension lacks the periodicity. Additionally, the model tends to prioritize the initial channels, causing a gradual loss of information in the later channels that correspond to the final stages of the simulation. Future research could address this limitation by exploring alternative strategies, such as decreasing the timestep between frames or mirroring techniques. This approach would enforce consistent learning across all channels, potentially improving accuracy throughout the simulation.

The initial training focused on the two-dimensional velocity field $\mathbf{u} = (u_x, u_y)$. The learning process was divided into two separate paths: one dedicated to $u_x$ and the other to $u_y$. Each path updated its weights independently, allowing the model to learn each velocity component autonomously. While the training employs distinct sets of weights for $u_x$ and $u_y$, the main goal is to analyze the overall behavior of the velocity field.

Our results show the model's ability to replicate the velocity field's magnitude by learning its two independent components. The FNO captures large and intermediate-scale features in the velocity field, consistent with spectral analysis. Visually, the predicted and target fields appear indistinguishable. However, a detailed evaluation using metrics such as MSE and SSIM reveals the presence of errors. These errors come from limitations in predicting small-scale features, whereas numerical errors are more pronounced. The model's performance is limited by the number of modes used by the FNO, restricting its ability to capture fine details on small scales or broader structures. Overcoming this limitation may require training the model on larger grids with an increased number of modes, although this would significantly raise the computational cost during training. Despite these challenges, the FNO captures the system's overall dynamics well, demonstrating its adaptability to variations in viscosity that were not part of the training data. The main limitation lies in predicting small-scale features, an issue also observed in numerical simulations \citep{Baerenzung2008, Parashar2010}. Error comparisons show a maximum deviation of $1.46 \times 10^{-1} $ and a minimum of $1.17 \times 10^{-9} $. While the FNO excels in modeling the system's general behavior, it remains constrained in accurately resolving small-scale details.

Moving on to the density field, it becomes evident that reproducing this field poses more significant challenges. Visually, the plots at $t_A \sim 1$ display no easily distinguishable characteristics. When analyzing the errors, they reach low values comparable to those observed in the vectorial fields. A remarkable difference surfaces when examining the density evolution. After each block prediction, parallel lines emerge, accompanied by enhanced errors. These parallel lines align with the peaks identified in the analysis of the kinetic and magnetic fields, correlating with the depth of the blocks. This suggests that deeper blocks require additional modes or a wider architectural design to mitigate the challenges.

Another goal is to evaluate the computational efficiency of our approach compared to traditional numerical techniques. Numerical codes are often time-consuming and computationally expensive, and many are not optimized for GPU usage. In recent years, there has been growing interest in using artificial intelligence to accelerate simulations. AI models can replicate entire simulations within seconds or minutes, but they are limited by the datasets they are trained on, whereas numerical codes can handle a wider range of problems. Our approach seeks to strike a balance by combining numerical simulations with AI models. By simulating just $20\%$ of the process using traditional methods, our model can accurately predict the remaining $80\%$, accelerating the overall simulation.

Future work should focus on addressing the limitations associated with small-scale features, potentially by training the model on larger grids and incorporating more Fourier modes into the architecture. Another promising direction is applying FNO to other traditional test cases, such as the Riemann problem, to further validate its performance across different scenarios and possibly accelerating the traditional approximate solvers, which
consume substantial computational time in finite-volume, shock-capturing codes. Additionally, exploring various astrophysical applications could uncover new opportunities for using FNO as a powerful tool in this field.

\section{Conclusions}

In this study, we explore the application of state-of-the-art architecture in the context of non-linear MHD simulations. The MHD regime is governed by partial differential equations describing the behavior of fluid density, velocity, and magnetic fields. The non-linear nature of these equations makes finding solutions a complex task. Traditional approaches involve employing time-consuming numerical solvers with high computational costs. ML-based models present a data-driven alternative for accelerating fluid simulations. The FNO stands out as an algorithm that operates within Fourier space, prompting an investigation into its potential using a well-established MHD test case known as the Orszag-Tang vortex. The Orszag-Tang vortex is a well-known problem that generates turbulence after a determined initial condition is given. The vortex is used as a problem to test numerical solvers, so naturally, testing FNO with the same test cases is a way to prove its efficiency. Our main conclusions can be summarized as follows.

(i) Effective Reproduction of Vector Fields: the FNO accurately learned and reproduced the velocity and magnetic fields in the Orszag-Tang vortex simulations, even in the presence of chaotic and turbulent behavior.

(ii) Accurate Learning of Scalar Fields: The model successfully captured the high variability in the fluid density field, contributing to a comprehensive understanding of density fluctuations.

(iii) Generalization to Unseen Physical Parameters: the FNO demonstrated robustness by accurately simulating scenarios with new combinations of physical parameters (for example, viscosity and diffusivity) that were not included in the training set.

(iv) Reproduction of Multi-scale Dynamics: The model effectively reproduced large and intermediate scales in the fluid system, as evidenced by the similarity in the power spectra for low and intermediate Fourier modes, the scales covered by the limited modes used in the FNO.

(v) Tracking Energy Dissipation Rates: FNO tracked the decrease in energy dissipation rates similarly to traditional numerical solvers, showcasing its effectiveness in capturing key physical behaviors despite some noted variations.

(vi) Significant Computational Speed-Up: The FNO achieved a speed-up of 25 times compared to traditional numerical methods, highlighting its potential as a tool for accelerating MHD simulations.

This work has several limitations associated with the prediction of small-scale features and the organization of data blocks. The limitations include: (a) the FNO struggles to accurately reproduce the depth of the blocks, (b) difficulties in replicating small-scale features such as dissipation rate and power spectrum at high values of $k$, (c) constraints on the temporal resolution due to the limited timestep in input and output data, (d) the two-dimensional nature of the data, which prevents the model from capturing three-dimensional aspects, and (e) the utilization of a limited number of Fourier modes in this study.

In conclusion, our investigation demonstrates the potential of FNO as a tool for simulating MHD flows. By successfully applying FNO to the well-known Orszag-Tang vortex, we showcase its ability to reproduce the magnetic and velocity vector fields. Despite limitations in predicting small-scale features, our findings highlight FNO's accuracy in capturing large and intermediate scales. Moreover, the model's capacity to generalize to never-seen combinations of physical parameters, such as viscosity and diffusivity, underscores its versatility. Perhaps most notably, FNO exhibits a speed-up, emphasizing its potential as a powerful tool to accelerate physical simulations. While challenges persist, our study suggests that FNO is promising as an efficient and effective surrogate for studying complex fluid systems.

% --------------------------------------------------------------------------------

\section*{Acknowledgments}

We acknowledge productive discussions with Ivan Almeida, Gustavo Soares, Clecio De Bom, Erik Almeida, Steven Farrell, Wahid Bhimji, Alejandro Cardenas-Avendano and Jonah Miller. R. D. acknowledges support by CAPES (Coordenação de Aperfeiçoamento de Pessoal de Nível Superior) Proex and Conselho Nacional de Desenvolvimento Cient\'ifico e Tecnol\'ogico (CNPQ). R. N. acknowledges support by: Fundação de Amparo à pesquisa do Estado de São Paulo (FAPESP) through grant 2022/10460-8, a Bolsa de Produtividade from CNPq and a NASA \textit{Fermi} Guest Investigator Grant (Cycle 16). We gratefully acknowledge the support of NVIDIA Corporation with the donation of two GPUs used for this research.

\section*{CRediT Authorship Contribution Statement}

\textbf{Roberta Duarte:} Conceptualization, methodology, software, investigation, writing -- original draft, visualization. \textbf{Rodrigo Nemmen:} Conceptualization, methodology, validation, writing -- review \& editing, supervision, project administration, funding acquisition. \textbf{Reinaldo Lima-Santos:} Conceptualization, methodology, Writing -- review \& editing.

\section*{Supplementary Materials}

A repository with the code used for training and inference will be made available soon. Likewise, for the training data.

\appendix

\section{}
\label{app:appendixA}

The values of viscosity and diffusivity used to generate data for training and validation is displayed in Figure \ref{numuappendix}. For each training session, the validation set is randomly selected from the full dataset, allowing the model to generalize across different fluid and magnetic field configurations. This approach ensures a robust evaluation of the model's performance while preventing overfitting to specific cases.

\section{}
\label{app:appendixB}

To test the robustness and flexibility of the FNO, we evaluated its performance under varying conditions not seen during training. Specifically, the model was tested on simulations with different grid resolutions and alternative parameter sets for $\nu$ and $\eta$, assessing its ability to generalize beyond the original training regime. Additionally, we investigated the impact of using a reduced number of Fourier modes, which directly affects the model's computational speed. These tests provide insight into the trade-off between performance and efficiency. In this section, we present results for a $64 \times 64$ grid resolution along with varying values of $\nu$ and $\eta$.

 \subsection{$\nu =  5 \times 10^{-4}$ and $\mu = 10^{-3}$}

\begin{figure*}[htbp]
 \centering
 \includegraphics[width=\textwidth]{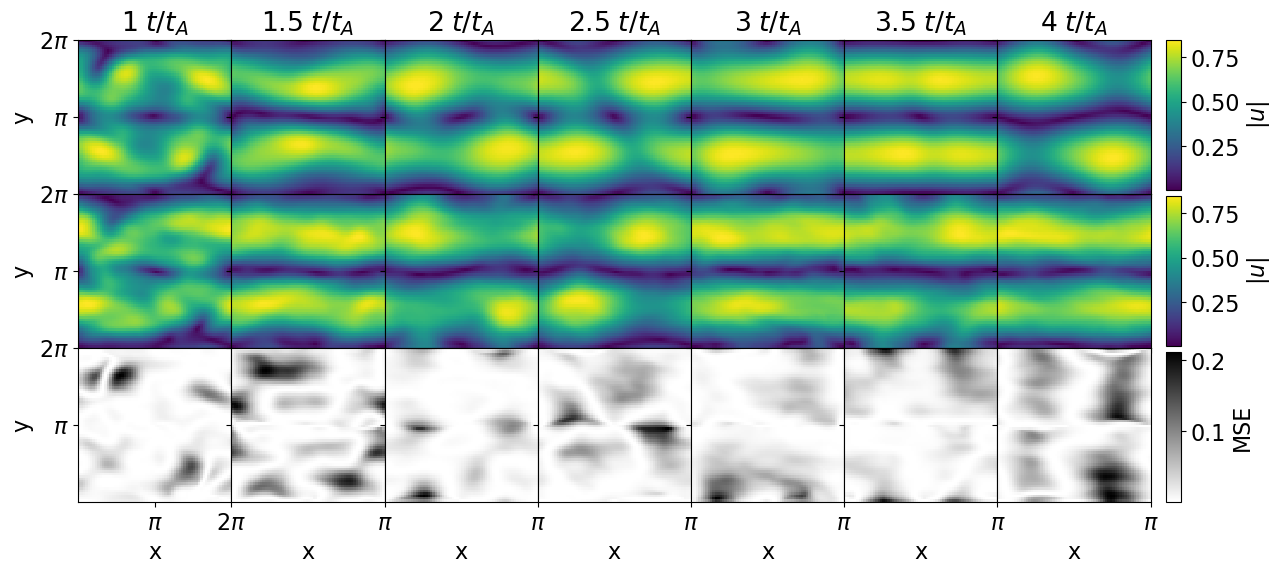}
    \caption{Comparison of the predicted and ground truth of $|\mathbf{u}|$ over 7 time steps from $t = 1$ to $t = 4\;t/t_A$. While the model accurately captures the large-scale dynamics in early stages, discrepancies increase over time.}
    \label{results:u_evolution_5e41e3}
\end{figure*}

\begin{figure*}[htbp]
 \centering
 \includegraphics[width=\textwidth]{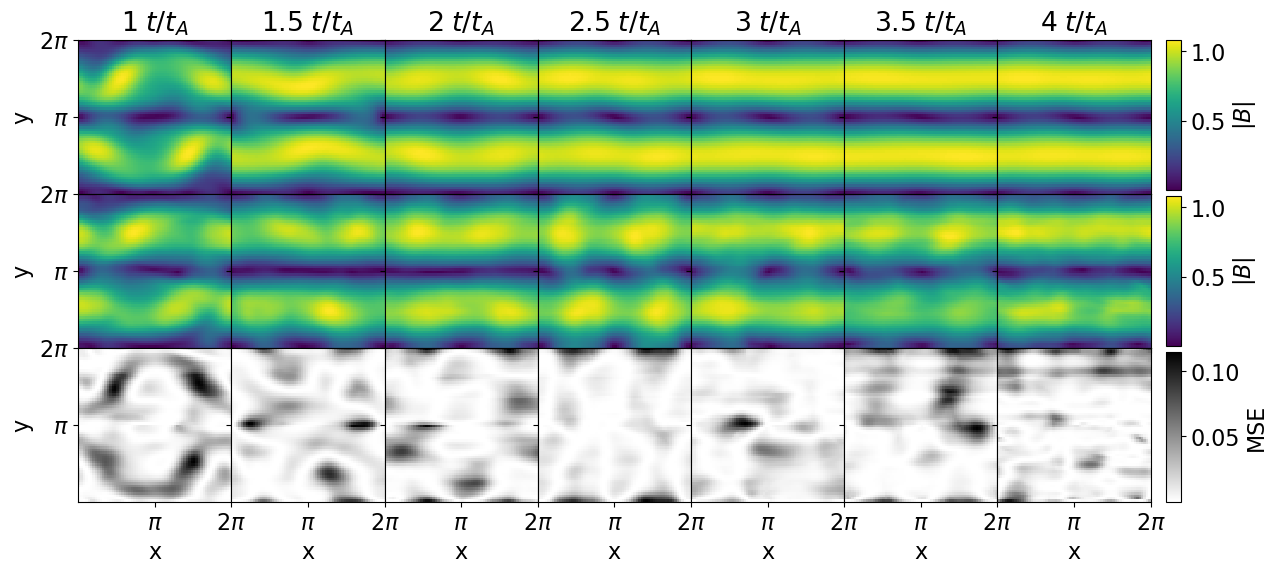}
    \caption{Same convention as Figure \ref{results:u_evolution_5e41e3}, the plot shows the result for $|\mathbf{B}|$ component evolution.}
    \label{results:b_evolution_5e41e3}
\end{figure*}

 Table \ref{appendixB:table} provides metrics of the predictions at $t = 1\; t/t_A$, presenting error metrics such as the mean, maximum, and minimum values for both velocity and magnetic field comparisons. In Figures \ref{results:u_evolution_5e41e3} and \ref{results:b_evolution_5e41e3}, we compare the predictions of $\mathbf{u}$ and $\mathbf{B}$ against the ground truth over 7 time steps. While the model initially captures the dynamics with reasonable accuracy, discrepancies begin to emerge in the later stages, where the predictions start to diverge from the target. This degradation in performance can be attributed to the lower spatial resolution ($64 \times 64$) combined with a limited number of Fourier modes (8), which restricts the model’s ability to represent finer-scale structures and long-term temporal dynamics. The increasing deviation suggests that higher variability and complex interactions present in later time steps are not fully captured under the current configuration. To address this, we propose increasing the number of Fourier modes to 16 or 32 during training, which may enhance the model’s robustness, especially for long-term forecasting.

\begin{table}[htbp]
\begin{center}
\begin{tabular}{|l|l|l|}
\hline
 & $|\mathbf{u}|$ &  $|\mathbf{B}|$ \\
\hline
Maximum Error & $2.11 \times 10^{-1}$ & $1.1 \times 10^{-1}$ \\ \hline
Minimum Error & $2.41 \times 10^{-11}$  & $1.89 \times 10^{-11}$ \\ \hline
Mean Error & $1.74 \times 10^{-2}$ & $2.15 \times 10^{-2}$ \\ \hline
Median & $5.32 \times 10^{-3}$ & $1.23 \times 10^{-2}$\\ \hline
SSIM Index & $0.64$ & $0.80$  \\ \hline
\end{tabular}
\caption{The table presents error metrics of $|\mathbf{u}|$ and $|\mathbf{B}|$ for the model's performance when $\nu = 5 \times 10^{-4}$ and $\mu = 1 \times 10^{-3}$.} 
\label{appendixB:table}
\end{center}
\end{table}

\begin{figure*}[htbp]
 \centering
 \includegraphics[width=\textwidth]{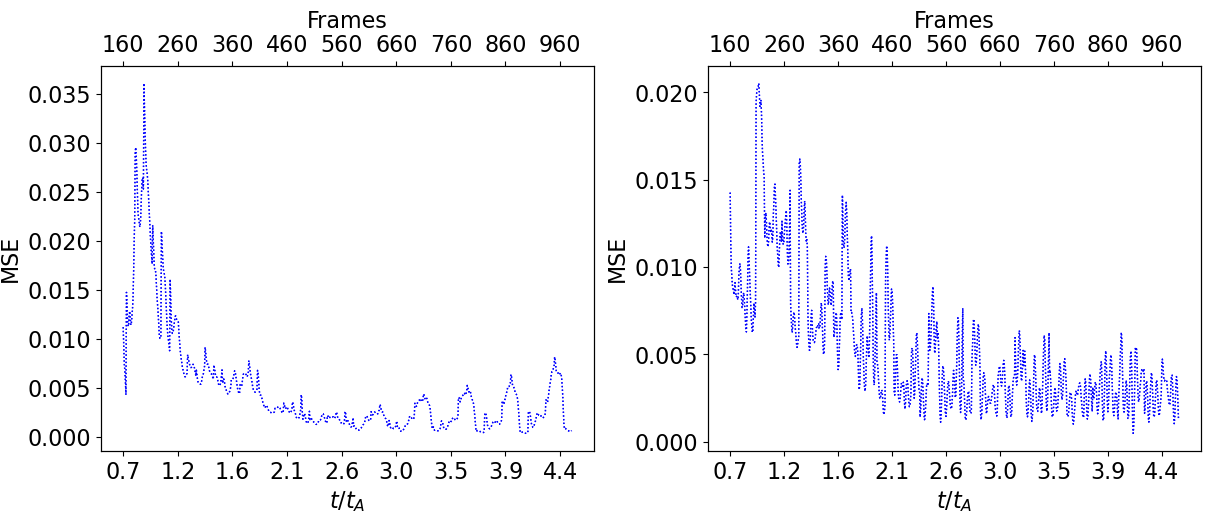}
    \caption{MSE over time between predictions and ground truth for $|\textbf{u}|$ and $|\textbf{B}|$ for the case $\nu = 5 \times 10^{-4}$ and $\mu = 10^{-3}$.}
    \label{results:error_5e41e3}
\end{figure*}

Figure \ref{results:error_5e41e3} shows the MSE between the predicted and ground truth for both $|\textbf{u}|$ and $|\textbf{B}|$. The left panel shows a pronounced initial error followed by a rapid decay and stabilization around $t = 1\;t/t_A$. This indicates that the model quickly adapts to the underlying velocity dynamics, achieving low prediction error at later times, despite some oscillations as presented in the Figure \ref{results:u_evolution_5e41e3}. In contrast, the right panel displays the MSE for $|\textbf{B}|$, which also follows a generally decreasing trend but with more fluctuations throughout the evolution. These persistent variations in the magnetic field error suggest a greater difficulty in capturing the complex dynamics likely due to small-scale structures and interactions.

 \subsection{$\nu =  3 \times 10^{-4}$ and $\mu = 10^{-4}$}

\begin{figure*}[htbp]
 \centering
 \includegraphics[width=\textwidth]{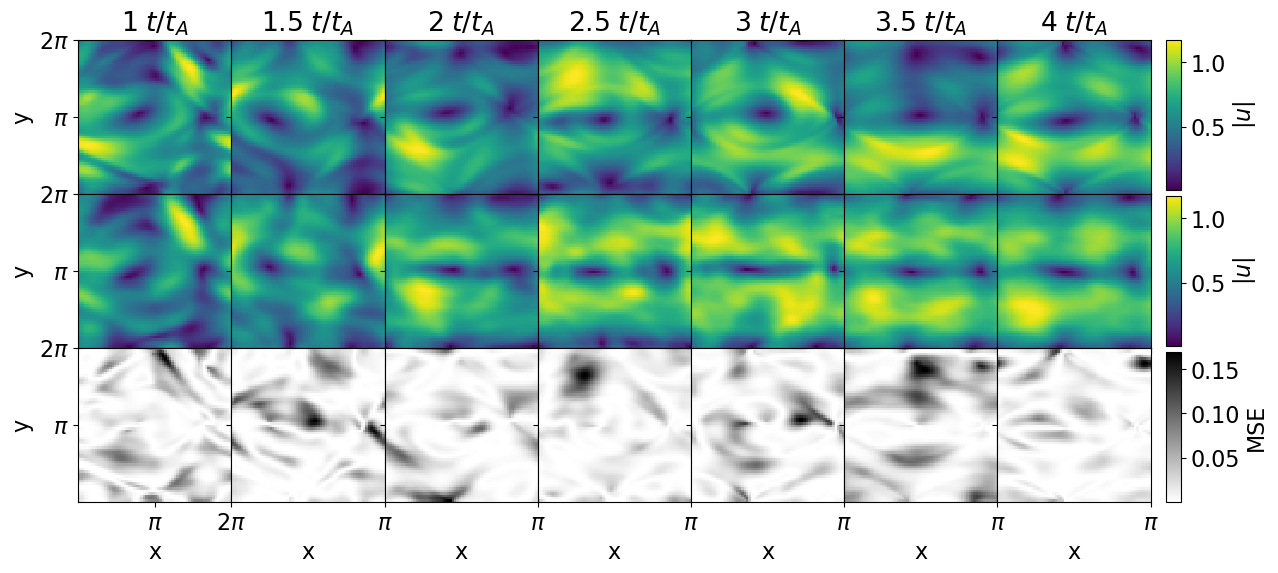}
    \caption{Evolution of $|\textbf{u}|$. Each column corresponds to a time step in Alfv\'en units. The first row displays the ground truth data while the second row shows predictions for a system with $\nu =  3 \times 10^{-4}$ and $\eta = 10^{-4}$. }
    \label{results:u_evolution_3e41e4}
\end{figure*}

\begin{figure*}[htbp]
 \centering
 \includegraphics[width=\textwidth]{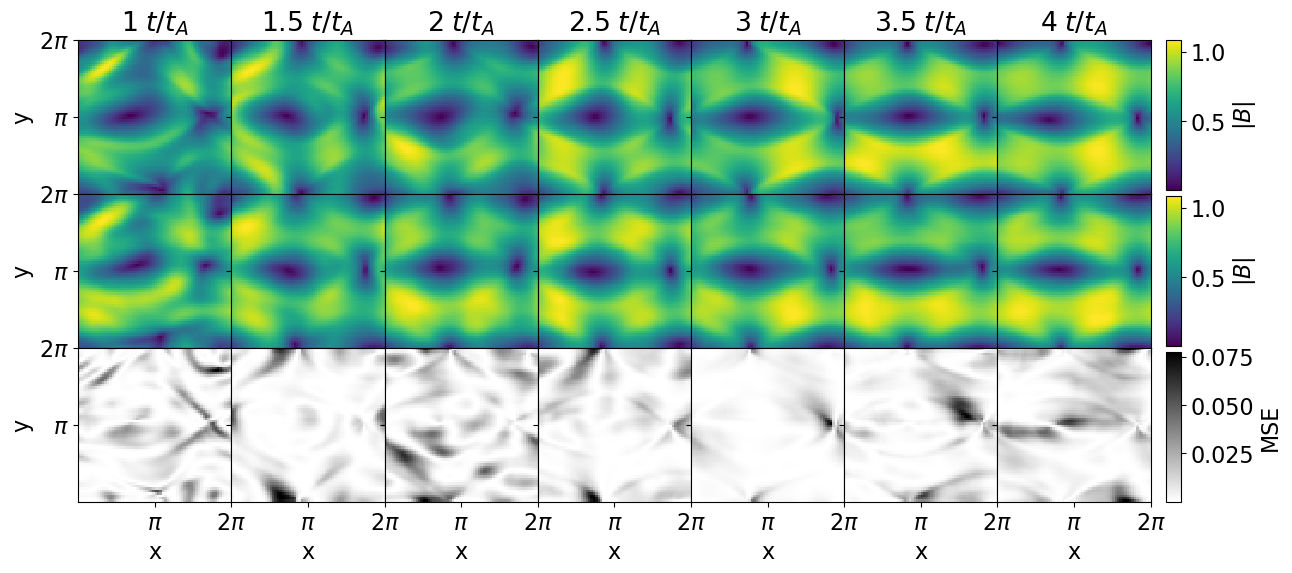}
    \caption{Same convention as Figure  \ref{results:u_evolution_3e41e4} but for $|\mathbf{B}|$.}
    \label{results:b_evolution_3e41e4}
\end{figure*}

 Here, we present the Table \ref{appendixB:table2} with the metrics results for $t = 1\;t/t_A$. Figure  \ref{results:u_evolution_3e41e4} presents the spatial-temporal evolution of $|\textbf{u}|$, with each column corresponding to a different time step $t/t_A$, ranging from 1 to 4 in increments of 0.5. The first row shows the ground truth data from MHD simulations, while the second row illustrates the predictions from the FNO. It was simulated with $\nu =  3 \times 10^{-4}$ and $\mu = 10^{-4}$, and trained with a spectral resolution limited to 8 Fourier modes. The third row displays the MSE between the predicted and target fields. Initially, the model captures the general fluid structures but as time progresses, especially in vortex-dominated regions, the error increases. It reveals the model's limitations in capturing 
 nonlinear interactions. Similarly, Figure \ref{results:b_evolution_3e41e4} shows a similar trend.

\begin{table}[h]
\begin{center}
\begin{tabular}{|l|l|l|}
\hline
 & $|\mathbf{u}|$ &  $|\mathbf{B}|$ \\
\hline
Maximum Error & $1.69 \times 10^{-1}$ & $7.75 \times 10^{-2}$ \\ \hline
Minimum Error & $9.20 \times 10^{-11}$  & $1.34 \times 10^{-11}$ \\ \hline
Mean Error & $1.92 \times 10^{-2}$ & $7 \times 10^{-3}$ \\ \hline
Median & $ 10^{-2}$ & $3.0 \times 10^{-3}$\\ \hline
SSIM Index & $0.67$ & $0.85$  \\ \hline
\end{tabular}
\caption{The table presents error metrics of $|\mathbf{u}|$ and $|\mathbf{B}|$ for the model's performance when $\nu = 3 \times 10^{-4}$ and $\mu = 1 \times 10^{-4}$.} 
\label{appendixB:table2}
\end{center}
\end{table}

\begin{figure*}[htbp]
 \centering
 \includegraphics[width=\textwidth]{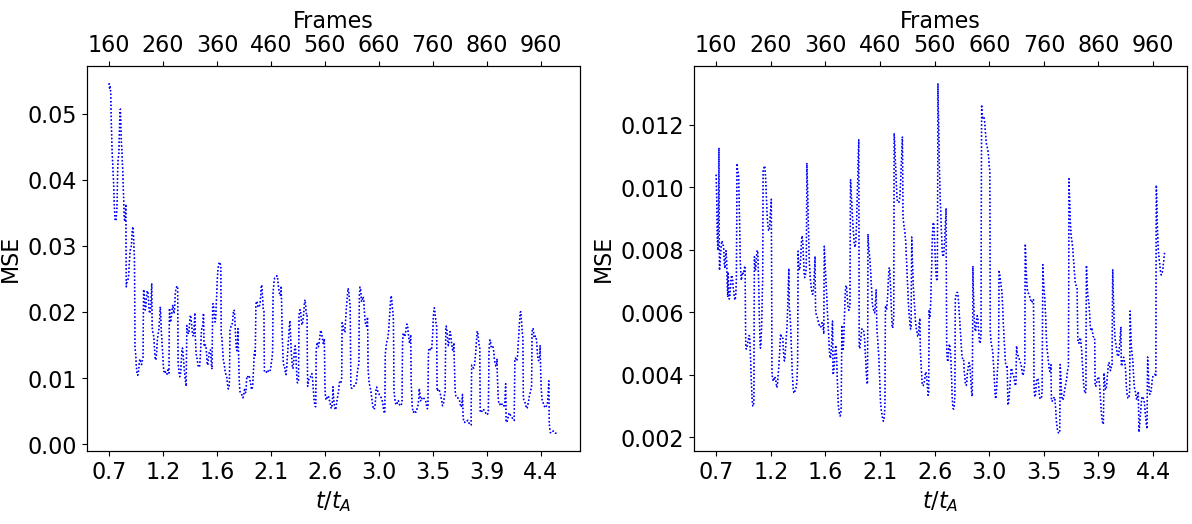}
    \caption{MSE over time between predictions and ground truth for $|\textbf{u}|$ and $|\textbf{B}|$ for the case $\nu = 3 \times 10^{-4}$ and $\mu = 10^{-4}$.}
    \label{results:error_3e41e4}
\end{figure*}

Similar to Figure \ref{results:error_5e41e3}, Figure \ref{results:error_3e41e4} shows the MSE between predicted and ground truth values under the parameter setting $\nu = 3 \times 10^{-4}$ and $\mu = 10^{-4}$. In the left panel, the MSE for $|\textbf{u}|$ starts with a relatively high value above 0.05 and decays quickly until around $t = 1\;t/t_A$, after which it exhibits a sustained oscillatory behavior with a downward trend, stabilizing around lower values for later times. This suggests the model captures the global velocity dynamics well but retains temporal variations likely tied to complex nonlinear structures. The right panel, showing the MSE for $|\textbf{B}|$ remains lower in magnitude overall and exhibits more frequent and sharper oscillations throughout the entire time range. Unlike the velocity field, the magnetic field error does not show a strong decay trend, indicating persistent difficulty in learning finer-scale magnetic features. Both results agree with the limitations associated with truncated Fourier modes that impact the learning of small structures. Also, in both cases, the lower resolution can be a factor as well.

\section{}
\label{app:appendixC}

In this Appendix, we will present the results for each component of the velocity and magnetic field when the $\nu = \mu = 5 \times 10^{-5}$. These results show how the models capture each component's evolution in the turbulent system. This in-depth analysis helps to further validate the FNO performance and offers insights into the challenges of forecasting different physical quantities in a highly nonlinear regime.

\subsection{Velocity Field Components}

\begin{figure*}[htbp]
 \centering
 \includegraphics[width=\textwidth]{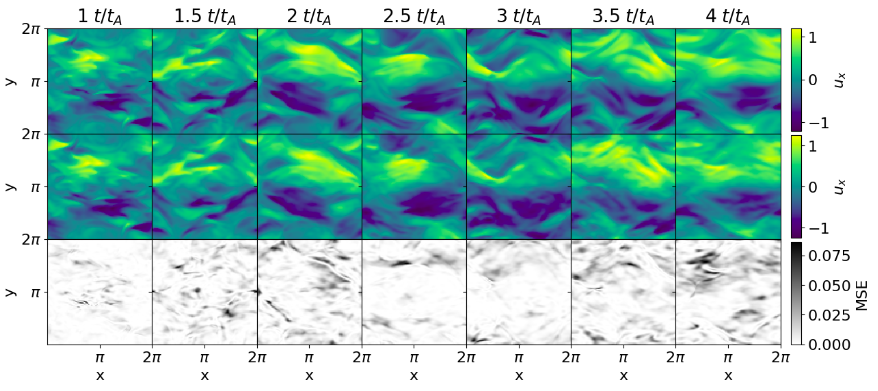}
    \caption{Temporal evolution of the $u_x$ velocity component from $t = 1$ to $t = 4\;t/t_A$. The plot shows the target, prediction and MSE in the first, second and third row respectively. The results demonstrate the fluid dynamics and how well the FNO successfully captures the turbulent flow.}
    \label{results:ux_evolution}
\end{figure*}

\begin{figure*}[htbp]
 \centering
 \includegraphics[width=\textwidth]{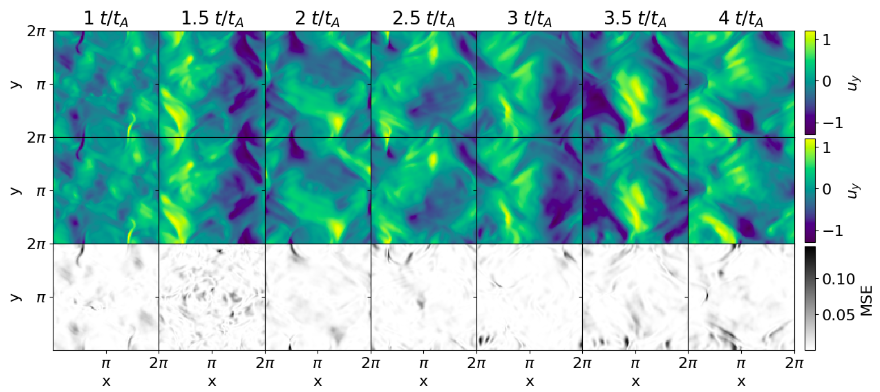}
    \caption{Similarly to Figure \ref{results:ux_evolution}, the plot shows the result for $u_y$ component evolution.}
    \label{results:uy_evolution}
\end{figure*}

Figure \ref{results:ux_evolution} presents the temporal evolution of $u_x$ component, showing that the model's predictions match the target, except in regions where there are sharp gradients or significant differences. The highest errors occur in areas with strong variations, indicating that the model struggles to capture regions with discontinuities. These results are consistent with the findings discussed in the \ref{results} section of this work. Similarly, Figure \ref{results:uy_evolution} displays the predictions for the $u_y$ component, where the same trend is observed. Both figures confirm that while the model effectively captures the general structure of the velocity field, it faces challenges in accurately resolving abrupt changes, reinforcing the conclusion that predicting discontinuities remains a limitation in this approach.

\subsection{Magnetic Field Components}

\begin{figure*}[htbp]
 \centering
 \includegraphics[width=\textwidth]{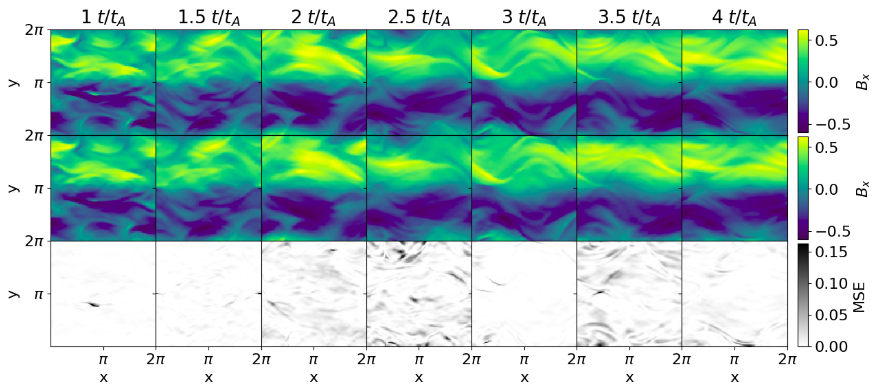}
    \caption{The temporal evolution of the $B_x$ $t = 1$ to $t = 4\;t/t_A$  is shown. The first row presents the target solution, the second row is the FNO predictions, and the third row is the MSE at each time step.}
    \label{results:bx_evolution}
\end{figure*}

\begin{figure*}[htbp]
 \centering
 \includegraphics[width=\textwidth]{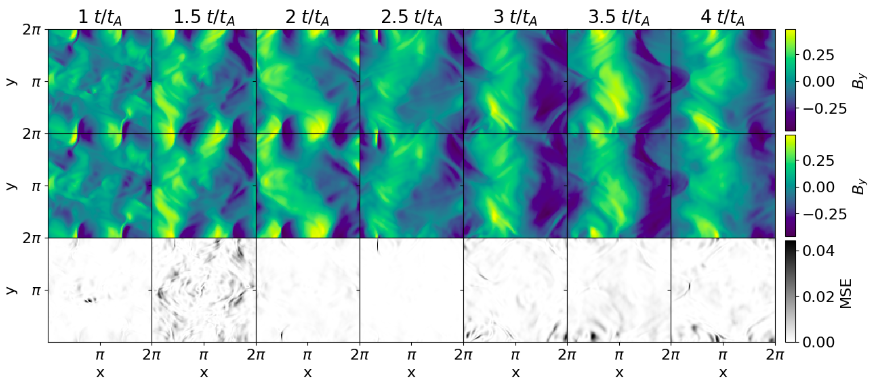}
    \caption{ This plot illustrates the evolution of the $B_y$ component, similar to the representation in Figure \ref{results:bx_evolution}.}
    \label{results:by_evolution}
\end{figure*}

Figures \ref{results:bx_evolution} and \ref{results:by_evolution} illustrate the evolution of the velocity components $B_x$ and $B_y$ over time. The predictions generally follow the target fields well, reproducing the large-scale flow structures. However, discrepancies become apparent in regions with rapid spatial changes, where the predicted values deviate. These differences point to a reduced accuracy in areas of steep gradients, which is a known limitation when using neural operators, particularly in the presence of discontinuities. The results underscore that, despite capturing the dominant flow patterns, the model still faces challenges in resolving fine-scale and abrupt variations within the velocity field due to truncation of Fourier modes.

\bibliographystyle{ieeetr}
\bibliography{main,refs,AI}

\end{document}